\documentclass{article}
\input epsf
\def\be{\begin{equation}}
\def\ee{\end{equation}}
\def\bea{\begin{eqnarray}}
\def\eea{\end{eqnarray}}
\def\phii{\phi_{i}}

\def\phihati{\hat{\phi_{i}}}
\def\phihatj{\hat{\phi_{j}}}
\def\Phii{\Phi_{i}}
\def\Phij{\Phi_{j}}
\def\Phihati{\hat{\Phi_{i}}}
\def\Phihatj{\hat{\Phi_{j}}}

\def\sof{\frac{\delta^{2} S}{\delta\Phi_i \delta\Phi_j}|_{\Phi=\phi}}
\def\dvdphi{\frac{\partial V}{\partial\phii}}
\def\vecx{\underline{x}}

\def\veck{\underline{k}}

\def\momvol{\frac{d^{3}\underline{k}}{(2\pi)^{3}}}
\def\nn{\nonumber}
\newcommand{\lrb}{\left(}
\newcommand{\rrb}{\right)}
\newcommand{\ba}{\begin{array}}
\newcommand{\ea}{\end{array}}

\def\adag-k{a^{\dag}_{-k}}
\def\a-k{a_{-k}}
\def\ak{a_{k}}

\def\ddag-k{d^{\dag}_{-k}}
\def\d-k{d_{-k}}
\def\dk{d_{k}}

\def\Bdagk{A^{\dag}_{\theta(k)}}
\def\Bdag-k{A^{\dag}_{\theta(k)}}
\def\B-k{A_{\theta(-k)}}
\def\Bk{A_{\theta(k)}}

\def\Ck{D_{\theta(k)}}
\def\Cdagk{D^{\dag}_{\theta(k)}}
\def\C-k{D_{\theta(-k)}}
\def\Cdag-k{D^{\dag}_{\theta(k)}}

\def\Fk{F_{k}}
\def\Fdagk{F^{\dag}_{k}}
\def\F-k{F_{k}}
\def\Fdag-k{F^{\dag}_{-k}}

\def\Adagk{A^{\dag}_{k}}
\def\Adag-k{A^{\dag}_{-k}}
\def\A-k{A_{-k}}
\def\Ak{A_{k}}

\def\Ddag-k{D^{\dag}_{-k}}
\def\Gdagk{G^{\dag}_{k}}
\def\Gdag-k{G^{\dag}_{-k}}
\def\G-k{G_{-k}}
\def\Gk{G_{k}}

\def\bmat{\lrb\begin{array}{c}}
\def\emat{\end{array}\rrb}
\def\phii{\phi_{i}}
\def\lb{\label}

\title{From QFT to DCC }
\author{{\bf B.Bambah\footnote{This paper is dedicated to Prof. Yoichiro Nambu for his $82^{nd}$ birthday.}}\\ School of Physics \\ University of
  Hyderabad, Hyderabad-500 046,India \\Email: bbsp@uohyd.ernet.in\\ \& \\ {\bf C.Mukku}\\
       Department of
  Mathematics \\and\\ Department of Computer Science \& Applications\\ Panjab University, Chandigarh-160014,India \\ Email: mukku@pu.ac.in, mukkuc@ic4life.net}
\begin{document}
\maketitle

\begin{abstract}
  A quantum field theoretical  model  for the
dynamics of the  disoriented chiral condensate is presented. A
unified approach to relate  the quantum field theory directly to
the formation, decay and signals of the DCC and its evolution is
taken. We use a background field analysis of the $O(4)$ sigma
 model keeping one-loop quantum corrections (quadratic order in the fluctuations).
An evolution  of the quantum fluctuations in an external,
expanding metric which simulates the expansion of the plasma, is
carried out.
 We examine, in detail, the amplification of the low momentum pion
 modes
with two competing effects, the expansion rate  of  the plasma and
the transition  rate of the vacuum configuration from a metastable
state into a stable state.We show the effect of DCC formation on
the multiplicity distributions and the Bose-Einstein correlations.
\end{abstract}

\large
\section{Introduction}
The mechanism of spontaneous symmetry breaking  \cite{nambu}
occupies an important role in particle physics, cosmology
\cite{gravsym} and condensed matter physics \cite{zurek}. The
understanding of this phenomenon has lead to cross fertilization
between these fields and  many new ideas have emerged \cite{boya}.
 When a symmetry is spontaneously broken by a
vacuum state, it is well known that thermal effects at equilibrium
can restore the broken symmetry \cite{kirznitz}. Typically, there
exists a critical temperature $T_c$ at which the effective
potential  as a function of some order parameter $\phi$ develops
an absolute minimum at $\phi=0$ and the system undergoes a phase
transition and relaxes to this minimum at higher temperature. A
manifestation of this, in a second order phase transition is that
at $T=T_c$ the order parameter fluctuates at all scales and long
wavelength oscillations occur \cite{Jackiw},\cite{mukku}. For
non-equilibrium situations, the dynamics of the transition of a
system from a symmetric to a broken symmetry state leads to the
formation of new structures and the generation of entropy in the
form of particle production\cite{kibble,vilenkin,zurek}.
 In
high energy physics, there are two well known examples of symmetry
breaking: electroweak symmetry-breaking and  chiral
symmetry-breaking in strong interactions. One of the main features
of QCD, the underlying theory of strong interactions, is the
spontaneous breaking of its approximate $SU(2)_L \times SU(2)_R$
chiral symmetry. Spontaneous breaking of this approximate symmetry
explains the very small pion masses. If the symmetry breaking were
exact, pions would be massless Nambu-Goldstone bosons. Another
consequence  of this symmetry breaking is the presence of a
non-vanishing quark condensate in the vacuum. It is believed that
at very high temperature a quark-gluon plasma is formed in which
chiral symmetry is restored, and much effort is being made to
explore such a phase transition by means of high-energy hadron or
heavy-ion collisions \cite{her}. The question is :  How do we test
the spontaneous symmetry breaking mechanism  in a non-equilibrium
process directly?
Can we create a suitable condition so that the
vacuum state is disturbed for a small region of space-time and
 observe
different excitations and domain structures in the vacuum?

Recently it has been conjectured that this maybe possible for the
chiral symmetry-breaking in strong interactions given the present
high energy physics facilities \cite{bj1,kw,R,dcc}.

To describe aspects of QCD related to this symmetry, it is
convenient to introduce an effective theory, to use a low energy effective
sigma model as a model for the QCD phase transition because it
respects the SU(2)$\times$SU(2) chiral symmetry of QCD with two
light flavours of quark and it contains a scalar field ($\Sigma$)
that has the same chiral properties as the quark condensate. The
$\Sigma$ field can thus be used to represent the order parameter
for the chiral phase transition. The  chiral four-vector of fields
$(\Sigma, \mbox{\boldmath $\pi$})$, where $\Sigma$ represents the
quark condensate and the three $\pi_i$ are the pion fields. In the
physical vacuum, $(\Sigma, \mbox{\boldmath $\pi$})$ points in the
$\Sigma$ direction. If the chiral symmetry were exact then there
would be a ``chiral circle" of states degenerate with this vacuum
state. In practice, the symmetry is explicitly broken by the
current quark masses and so there is a unique vacuum.

Because of this circle of nearly degenerate field configurations,
as the chirally restored plasma cools and returns to the normal
phase, the system could form regions in which the chiral fields are
misaligned, that is, chirally rotated from their usual orientation
along the $\Sigma$ direction. There has been much recent interest
in this phenomenon, which is known as a disoriented chiral
condensate (DCC). If such a state were formed, it would lead to
anomalously large event-by-event fluctuations in the ratio of
charged to neutral pions.

A region of DCC can  be thought of as {\it a cluster of pions of
near identical momentum around zero( coherently produced) with an
anomalously large amount of fluctuation of the neutral fraction}.
In order to produce such a state in a quark gluon plasma, the hot
plasma must evolve far from equilibrium and in particular it must
reach an unstable configuration such that the long-wavelength pion
modes are amplified exponentially when the system relaxes to the
stable vacuum state. Thus questions of whether a DCC forms and how
it evolves cannot be addressed in the framework of equilibrium
thermodynamics. Techniques for applying QCD directly to such
situations do not exist at present. To explain these
non-equilibrium phenomena, we need to restructure the theory of
phase transitions to incorporate the micro structures (or states)
instead of macro structures. For this we need to set up a theory
at the quantum level. Some attempts have been made
\cite{KK},\cite{Gavin},\cite{cooper},\cite{lampert},\cite{randrup},
but to our knowledge these are incomplete as they do not
incorporate the disorientation of the condensate and the evolution
of the plasma in one unified picture. In his talk at Trento,
\cite{bj3} Bjorken pointed out some of the unresolved issues
concerning the DCC called the {\it DCC trouble list}. They
were:\begin{itemize}\item (a) Are coherent states the right
quantum DCC description, or should one go beyond to squeezed
states, etc.
\item (b) How does one link DCC thinking
to quantum effects in the data, especially the Bose­Einstein
correlations. Is DCC just another way of talking about the same
thing?
\item(c) How does one enforce quantum number conservation,
especially charge? \cite{suzuki}
\item(d) DCC production may imply anomalous bremsstrahlung, due to
the large quantum fluctuations in charge. Can this be calculated
from first principles?\item (e) Can one really set up the problem
at the quantum field theory level?\end{itemize}

In this paper we lay the foundation for a unified quantum picture
of the DCC with and without orientation and give our answers to
these questions.
\section{The Model}
In most treatments of DCC formation, the classic Gellman-Levy
lagrangian  \cite{gell} is used. For a more general treatment
which encompasses this Lagrangian, we use the field theoretic
$O(4)$ symmetric sigma model with Lagrangian density \be{\it L}=
(\frac{1}{2} \dot{\Phi}_{i}^{2}
-\frac{1}{2a^{2}}(\nabla\Phi_i)^{2} -\frac{1}{2} m^{2}\Phi_{i}^{2}
-\frac{\lambda}{4\!}(\Phi_{i}^{2})^2 )  \lb{o4}\ee \be \Phi_i=
 \lrb \ba{c}\Phi_1 \\  \Phi_2
\\  \Phi_3 \\ \Phi_4 \ea \rrb
\ee
 where $\Phi_i$, $i=1..4$ are real scalar fields.

We note here that the potential we use is not the one
traditionally used by the early practitioners of DCC analyses.
Thus, for later comparison  with the dynamical equations for the
pions used as starting points for analysing the DCC,  we will
later relate the potential that we have chosen to the traditional
potential in the Gell-Mann, Levy model \cite{Amado}. The
Lagrangian density in the Gell-Mann, Levy model is given by
 \be L={1\over2}\, (
\partial_\mu {\vec \pi}\, \partial ^{\mu} {\vec \pi} + \partial_\mu
\,\Sigma \partial^\mu \Sigma)- {\lambda\over 4}\, ( {\vec \pi}^2 +
\Sigma^2 -v^2)^2+H\Sigma. \ee The action is
\be
 S=\int d^{4}x
\left[{1\over2}(\partial_\mu\phi^a\partial_{\mu}\phi_{a}
-{\lambda\over4}(\phi^a\phi_a-v^2)^2+H\Sigma \right] .
\end{equation}
The basic object of study is then the chiral field
\begin{equation}
\vec{\phi}(r,t)=\Sigma(r,t)+ i\bf{\tau}\cdot\vec{\pi}(r,t),
\end{equation}
 where $\Sigma(r,t)$ and
 $\vec{\pi}(r,t)$ are  the  scalar and vector fields respectively
 of
the $O(4)$ vector $\vec{\phi}=(\Sigma,\vec{\pi})$and $\tau$ are
the Pauli matrices.\\

To make contact with the physical constants $\lambda$, $v$ and $H$
are related to the physical quantities, $f_\pi=92MeV$  (pion form
factor) and
 $m_\pi=138MeV/c^2$  by the relations
\begin{eqnarray}
m_{\pi}^2=\frac{H}{f_{\pi}}&=&\lambda(f_{\pi}^2-v^2)\\ \nn
m_{\Sigma}^2&=&3\lambda f_{\pi}^2-\lambda v^2\approx 2 \lambda
f_{\pi}^2=600 MeV/c^2 \\ \nn \lambda &=& {m_\Sigma^2c^2-m_\pi^2c^2
\over 2f_\pi^2}=20.14,\\ \nn
<v> &=&\left[{m_\Sigma^2-3m_\pi^2\over m_\Sigma^2-m_\pi^2}f_\pi^2\right]^{1\over2}
=86.71MeV\, \\ H &=& m_\pi^2c^2 f_\pi= (120.55MeV)^3.
\end{eqnarray}

Note that the fermionic part is neglected here, since the focus is
on the condensate formed in the symmetry broken phase where quark
degrees of freedom are already confined. With the
transformation,$\Sigma->\Sigma-v$,the minimum  of potential is at
$<\phi^2>=<\Sigma^2+\pi^2>=<v>^2=f_{\pi}^{2}$ and the usual
equilibrium vacuum  state is an ordered state
$<\Sigma>=<v>=f_{\pi}$ and  $<\pi>=0$.

Now consider the dynamic evolution of the system in a hot quark
gluon plasma. When the temperature $T \ge T_c$ the system reaches
a state of restored symmetry $<\phi^2>=0$. If the subsequent
expansion of the plasma is adiabatic, the$<\phi>$ field gradually
relaxes to the equilibrium state  as the system cools to below
$T_c$. This is called the "annealing" or adiabatic scenario. It
was first pointed out in reference \cite{kw} that if the cooling
process is very rapid and the system is out of equilibrium, i.e.,
in the event of a sudden quench from a state of restored symmetry
to a state of broken symmetry such as that occurring in   a
rapidly cooling expanding plasma, the configuration of the $\phi$
field will lag behind the expansion of the plasma and there is a
mismatch of the configurations and their evolution. After a
quench, the high temperature configuration does not have time to
react to the sudden change of the environment, thus  the vacuum
expectation value $<\phi>$ would stay
 what it was at high temperature for a while and then relax to its equilibrium value. This results in the
 formation of DCC domains and is known as the {\it Baked Alaska} \cite{bj} scenario.

Another non-equilibrium  situation that can arise  is  one where
the  system can go through a meta-stable disordered vacuum
 $<\Sigma>=f_{\pi}Cos(\theta)$ $<\pi>=f_{\pi}\overline{n}Sin(\theta)$
and then relax by quantum fluctuations to an equilibrium
configuration. Here $\theta$ measures the degree of disorientation
of the condensate in isospin space.

In both these non-equilibrium situations, the canonical approach
to the full quantum evolution of the fields is extremely difficult
to carry out explicitly. However, we may study the quantum
evolution of the mean field  and include fluctuations. In
calculating the effective Hamiltonian we use the $O(4)$ symmetric
linear sigma model, whose Lagrangian density is given by
(\ref{o4}). It is easy to see that with the identification
$\Phi^2=(\Sigma^2+\pi^2-f_{\pi}^2)$, and, $\Sigma->\Sigma-v$, the
effective potential in the O(4) linear sigma model  is  the same
as the Gell-Mann Levy model. We will use this identification when
considering the evolution of the pion field.

 The choice
of metric is dictated by the simple spherical geometry of the
problem. Assuming a homogeneous expansion of the plasma, a
Robertson-Walker type of metric is chosen with an expansion rate
$a(t)$. This will allow us to examine the competing rates leading
to ``rolling down rate'' of the vacuum and the expansion rate
$a(t)$. The line element describing the expansion of the plasma
bubble is chosen to be:
\be
ds^{2}=dt^{2}-a(t)^{2} d\vec{x}^{2}, \ee where $a(t)$ is the
expansion parameter.

The action of the $O(4)$ sigma model in this metric becomes:
\be
S=\int d^{3}x dt a(t)^{3} (\frac{1}{2} \dot{\Phi}_{i}^{2}
-\frac{1}{2a^{2}}(\nabla\Phi_i)^{2} -\frac{1}{2} m^{2}\Phi_{i}^{2}
-\frac{\lambda}{4\!}(\Phi_{i}^{2})^2 ), \ee with
 \be \Phi_i=
 \lrb \ba{c}\Phi_1 \\  \Phi_2
\\  \Phi_3 \\ \Phi_4 \ea \rrb
\ee
 where $\Phi_i$, $i=1..4$ are real scalar fields.

We now use a background field analysis to study the quantum effects.
Assume
$\Phii$ has a background classical component $\phii$ which
satisfies the classical equations of motion:
\be
\frac{\delta S}{\delta\Phii}|_{\Phii=\phii} = 0. \ee Treat the
quantum field $\phihati$ as  fluctuation around classical solution:
\be
\Phi_{i} \longrightarrow \phii + \phihati \ee
since  $\phii$ satisfies the classical equations of motion,
\be
S=S[\phii ] + \frac{1}{2}\phihati\sof\phihatj +\cdots. \ee

We shall restrict our analysis to quadratic fluctuations only. In
addition, we shall also drop the term $S[\phii]$ as it is just a
constant additive term to the quadratic action. Therefore we shall
deal with a quadratic fluctuation action given simply by:
\be
S_{2}=\frac{1}{2}\phihati\sof\phihatj . \ee For the particular
scalar field action given above, assuming all fields vanish at
infinity,
\be
\frac{\delta S}{\delta\Phij} = -\partial_{\mu}(a^3
g^{\mu\nu}\partial_{\nu}\Phij)-a^3 m^2\Phij -a^3 \frac{\delta
V}{\delta\Phij}. \ee Imposing the classical equations of motion, we
find that
\be
\partial_{\mu}(a^3 g^{\mu\nu}\partial_{\nu}\phii)+a^3 m^2\phii +a^3 \frac{\partial V}{\partial\phii} = 0
\ee where $\frac{\partial V}{\partial\phii}\equiv\frac{\partial
V}{\partial\Phii}|_{\phii}$.

The equations of motion in this  metric are
\be
3\frac{\dot{a}}{a}\dot{\phii} +\dot{\phii}^{2}
-\frac{1}{a^{2}}\underline{\nabla}^{2}\phii +m^{2}\phii +\dvdphi =
0. \ee Since we are interested in the dynamics of the fluctuation
field, we shall treat the fluctuation field in $S_{2}$ as a
classical field and $S_{2}$ itself as the classical action for its
dynamics. While the quadratic part in the fluctuations is reduced
to:
\be
S_{2}=\int d^{3}\underline{x} dt \frac{a^{3}}{2}
(\dot{\phihati^{2}}
-\frac{1}{a^{2}}(\underline{\nabla}\phihati)^{2} -
m^{2}\phihati^{2}-\phihati \frac{\delta^{2}
V}{\delta\Phihati\delta\Phihatj}|_{\phi}\phihatj). \ee We can
define a Lagrangian density for studying the dynamics of the
fluctuations, $\cal{L}$, as follows:
\be
{\cal{L}}=\frac{a^3}{2}(\phihati^{2}-\frac{1}{a^{2}}(\underline{\nabla}\phihati)^{2}
-m^{2}\phihati^{2} -\phihati\frac{\partial^{2}
V}{\partial\Phii\partial\Phij}|_{\phi}\phihatj). \ee Carrying out
a Legendre transformation, it is easy to write down the
Hamiltonian density
\be
{\cal{H}}={\frac{1}{2a^{3}}\hat{p_{i}}^{2}
+\frac{a}{2}(\underline{\nabla}\phihati)^{2}+\frac{a^{3}m^{2}}{2}\phihati^{2}
 +\frac{a^3}{2}(\phihati\frac{\partial^{2}
V}{\partial\Phii\partial\Phij}|_{\phi}\phihatj)} \ee where
\be
\hat{p_{i}}=\frac{\delta \it{L}}{\delta
\dot{\phihati}}=a^{3}\dot{\phihati}. \ee
We also have
\begin{equation}
\frac{\partial^2 V}{\partial\Phi_i \partial\Phi_j}|_\phi =2
\lambda \phi_i\phi_j+\lambda\phi_k^2 \delta_{ij}.
\end{equation}
Assume that the fluctuation field $\Phi_i$ decomposes into its
constituents as:\\
\begin{equation}
\hat{\phi}=<\Phi>-\phi=\left(\begin{array}{c}\pi_1\\ \pi_2\\
\pi_0\\ \Sigma\end{array}\right).
\end{equation}
We have started with an $O(4)$ quartet of scalar fields in order
that we can construct the dynamics of quenched pions in the
formation of a disoriented chiral condensate.
The physical fields are defined so that
\begin{equation}
  \pi_+=\frac{1}{\sqrt{2}}(\pi_1+i\pi_2);\hspace{2cm}
  \pi_-=\frac{1}{\sqrt{2}}(\pi_1-i\pi_2).
\end{equation}
Analogously, we define the classical background fields as:
\begin{equation}
  v_+=\frac{1}{\sqrt{2}}(v_1+iv_2);\hspace{2cm}
  v_-=\frac{1}{\sqrt{2}}(v_1-iv_2),\\
\end{equation}
following the identification:
\begin{equation}
\phi=\left(\begin{array}{c}v_1\\ v_2\\ v_3=v\\
\sigma\end{array}\right)\equiv <\Phi>.
\end{equation}
It is easy to see that the action takes the following form:
\begin{eqnarray}
S&=&\int d^{3}\vec{x} dt a^{3}\{
\dot{\pi}_{+}\dot{\pi}_{-}-\frac{1}{a^2}(\nabla\pi_{+})(\nabla\pi_{-})
-(m^{2}+(4\lambda)v_{+}v_{-}+\lambda
v_{3}^{2}+\lambda\sigma^{2})\pi_{+}\pi_{-}  \nn \\
&+&\frac{1}{2}\dot{\pi}_{0}^{2}-\frac{1}{2 a^{2}}(\nabla
\pi_{0})^{2}-\frac{1}{2}(m^{2}+(2\lambda)v_{+}v_{-}+3\lambda
v_{3}^{2}+\lambda\sigma^{2})\pi_{0}^{2}
+\frac{1}{2}\dot{\Sigma}^{2}-\frac{1}{2
a^{2}}(\nabla\Sigma)^{2}\nn \\
&-&\frac{1}{2}(m^{2}+(2\lambda)v_{+}v_{-}+3\lambda \sigma^{2}
 +\lambda v_{3}^{2})\Sigma^{2}
-\lambda(v_{-}^{2}\pi_{+}^{2}+v_{+}^{2}\pi_{-}^{2} +2\sigma
v_{3}\pi_{0}\Sigma
\nn \\
&+&2v_{-}v_{3}\pi_{0}\pi_{+}+2v_{+}v_{3}\pi_{0}\pi_{-}+2v_{-}\sigma\pi_{+}\Sigma+2v_{+}\sigma\pi_{-}\Sigma)\}
\end{eqnarray}
The Hamiltonian for this action can be written as\be
H=H_{neutral}+H_{charged}+H_{mixed}, \ee where \bea
H_{neutral}&=&\int d^3xdta^3 \{
\frac{(p_{0\pi_0}^2)}{2a^6}+\frac{1}{2a^2}(\nabla\pi_0^2)+\frac{1}{2}(m_{\pi}^2)(\pi_0)^2+\frac{(\Omega_{\pi}^2-\omega_{\pi}^2)}{2a^6}\pi_0^2
\nn \\
&+&\frac{(p_{0\Sigma}^2)}{2a^6}+\frac{1}{2a^2}(\nabla\Sigma^2)+\frac{1}{2}(m_{\Sigma}^2)(\Sigma)^2+\frac{1}{2a^6}(\Omega_{\Sigma}^2-\omega_{\Sigma}^2)(\Sigma)^2\} \eea
\bea
H_{charged}&=&\int d^3x
dt a^3 (\frac{1}{a^6} \{(p_{0+}p_{0-})\nn \\
&+&\frac{1}{a^2}(\nabla\pi_-)(\nabla\pi_+)+(m_\pi^2)(\pi_+\pi_-)+\frac{1}{a^6}(\Omega_{\pi_{\pm}}^2-\omega_{\pi_{\pm}}^2)(\pi_+\pi_-)\}
\eea \bea
 H_{mixed}&=&\int d^3x
a^3dt\{\lambda(v_{+}^2\pi_{-}^2+v_{-}^{2}\pi_{+}^2)
 \nn \\
&+&(2\lambda)(v_+v_3\pi_-\pi_0+ v_-v_3\pi_+\pi_0+\sigma
v_3\Sigma\pi_0+v_+\sigma\pi_-\Sigma+\sigma v_-\Sigma\pi_+)\}.
\eea Where we have also put \bea
\frac{\Omega^2_{\pi_0}-\omega^2_{\pi_0}}{a^6}&=&\lambda[v^2+2v_3^2]
\nn \\ \frac{\Omega^2_{\pi
{\pm}}-\omega^2_{\pi_{\pm}}}{a^6}&=&\lambda[v^2+2v_+v_-]
\nn \\
\frac{\Omega^2_{\Sigma}-\omega^2_{\Sigma}}{a^6}&=&\lambda[v^2+2\sigma^2]
\eea and
\be
2v_+v_-+v_3^2+\sigma^2=v^2. \ee This is the most general
Hamiltonian for the pion sigma system in the background field
formalism. The background field can now be parametrized through three angles:\be
\phii=\lrb \ba{c}
vCos(\rho)Sin(\theta)Sin(\alpha)\\vCos(\rho)Sin(\theta)Cos(\alpha)\\vSin(\rho)Sin(\theta)\\vCos(\theta)\ea
\rrb.\ee
In order to consider all the special cases that are
possible in a transparent way, we shall simplify the parametrization
of the possible form for the background field to two angles,
$\theta$ and $\rho$ by letting $\alpha=\frac{\pi}{4}$:
then,
$v\pm=\frac{v}{\sqrt{2}}Cos(\rho)Sin(\theta)$,\hspace{0.5cm}$v_{3}=v
Sin(\rho)Sin(\theta)$,\hspace{0.5cm} and
\hspace{0.5cm}$\sigma=vCos(\theta)$.

\section{Quantisation}
 Using standard Canonical quantisation techniques the mode decomposition of the
Hamiltonian is:
 \bea
 H_{neutral}=\int \momvol
\frac{1}{2}\{
\frac{\omega_{\pi}}{a^3}(a_k^{\dag}a_k+a_ka^{\dag}_k)  \nn \\
+\frac{\omega_{\pi}}{2a^3}(\frac{\Omega_{\pi}^2}{\omega_{\pi}^2}-1)(a_k^{\dag}a_k+a_ka_k^{\dag}+a_{-k}a_{k}+a_{-k}^{\dag}a_{k}^{\dag})
 \nn \\
 +\frac{\omega_{\Sigma}}{a^3}(
 d_k^{\dag}d_k+d_kd^{\dag}_k)+\frac{\omega_{\Sigma}}{2a^3}(\frac{\Omega_\Sigma^2}{\omega_{\Sigma}^2}-1)(d_k^{\dag}d_k+d_kd_k^{\dag}+d_{-k}d_{k}+d_{-k}^{\dag}d_{k}^{\dag})\}
\eea
\bea
 H_{charged}=\int \momvol
\{ \frac{\omega_{\pi}}{a^3}(b_k^{\dag}b_k+c_kc^{\dag}_k)  \nn \\
+\frac{\omega_{\pi}}{2a^3}(\frac{\Omega_{\pi_\pm}^2}{\omega_{\pi}^2}-1)(b_k^{\dag}b_k+c_kc_k^{\dag}+b_{-k}c_{k}+c_{-k}^{\dag}b_{k}^{\dag})\}
\eea
\bea
 H_{mixed}=\int \momvol
\{ \frac{\lambda a^3v^2cos^2(\rho)sin^2(\theta)}{4\omega_\pi}(
 b_k b_{-k}+b_{k}c^{\dag}_k
 \nn \\+c^{\dag}_k b_{k}+c_k c_{-k}+c^{\dag}_k
 c^{\dag}_{-k}+c_kb^{\dag}_k+b^{\dag}_{k}c_{k}+b^{\dag}_{k}b^{\dag}_{-k})
 \nn \\
+ \frac{\lambda
a^3v^2cos(\rho)sin(\rho)sin^2(\theta)}{\sqrt{\omega_{\Sigma}\omega_\pi}}
\nn\\
\lrb b_k a_{-k}+b_{k}a^{\dag}_k+c^{\dag}_k a_{k}+c_k
a_{-k}+c^{\dag}_k
a^{\dag}_{-k}+c_ka^{\dag}_k+b^{\dag}_{k}a_{k}+b^{\dag}_{k}a^{\dag}_{-k}\rrb \nn \\
+ \frac{\lambda
a^3v^2sin(\rho)sin(\theta)cos(\theta)}{\sqrt{\omega_\pi\omega_{\Sigma}}}\lrb
 d_k a_{-k}+d_{k}a^{\dag}_k+d^{\dag}_k a_{k}+d^{\dag}_k
 a^{\dag}_{-k}\rrb
 \nn \\
+ \frac{\lambda
a^3v^2cos(\rho)sin(\theta)cos(\theta)}{\sqrt{\omega_{\pi}\omega_{\Sigma}}}
\nn \\ \lrb
 b_k d_{-k}+b_{k}d^{\dag}_k+c^{\dag}_k d_{k}+c_k d_{-k}+c^{\dag}_k d^{\dag}_{-k}+c_kd^{\dag}_k+b^{\dag}_{k}d_{k}+b^{\dag}_{k}d^{\dag}_{-k}\rrb
 \}\eea
 where \bea
\frac{\omega^{2}_{\pi}(k)}{a^{6}}\equiv\frac{\omega^{2}_{\pi_0}(k)}{a^{6}}=\frac{\omega^{2}_{\pi_\pm}(k)}{a^{6}}=(m_{\pi}^{2}+\frac{\veck^{2}}{a^{2}})\nn\\
\frac{\omega^{2}_{\Sigma}(k)}{a^{6}}=(m_{\Sigma}^{2}+\frac{\veck^{2}}{a^{2}})
\eea and \be
\frac{\Omega_{\pi}^{2}(k)}{a^6}=\frac{k^2}{a^2}+m_{\pi}^{2}+\lambda
(v^{2}+2v_{3}^{2}).\ee
\be
\frac{\Omega_{\pi_\pm}^{2}(k)}{a^6}=\frac{k^2}{a^2}+m_{\pi_\pm}^{2}+\lambda(v^{2}+2v_{+}v_{-}).\ee

An important point to note here is that although
$\Omega_{\pi}(k)$ and $\omega_{\pi}(k)$ are momentum dependent
quantities, for ease of notation we will drop the k dependence for
further calculations in this section and revive it when necessary
for the description of the physical processes.

It is very interesting to note that if either of $v_{\pm},v_3$ is
zero, then we obtain the usual expected dynamics for the  pions
with back to back correlations. But if we allow for either
$v_{\pm}$ or $v_3$ to be non-zero as may be envisaged in the
highly non-equilibrium dynamics involving the formation of a
metastable DCC, then, we have terms which involve mixed
interactions of the pions and sigma.

Clearly, there are two interesting cases to be considered here: $v_{\pm} =0$ and $v_3=0$ (equivalently, $\theta=0$) and
the second case occurs with the formation of a metastable,
misaligned vacuum when $\rho=\frac{\pi}{2}$.

\noindent {\bf \underline{CASE 1}}

For this case we use  $v_{\pm} =0$ and $v_3=0$ (or, equivalently, $\theta=0$) showing that
 $H$ reduces to \bea
 H=\int \momvol
\{ \frac{\omega_{\pi}}{2a^3}(a_k^{\dag}a_k+a_ka^{\dag}_k) \nn \\
+\frac{\omega_{\pi}}{4a^3}(\frac{\Omega_{\pi}^2}{\omega_{\pi}^2}-1)(a_k^{\dag}a_k+a_ka_k^{\dag}+a_{-k}a_{k}+a_{-k}^{\dag}a_{k}^{\dag})
+\frac{\omega_{\Sigma}}{2a^3}(
 d_k^{\dag}d_k+d_kd^{\dag}_k)
\nn \\+\frac{\omega_{\Sigma}}{4
 a^3}(\frac{\Omega_\Sigma^2}{\omega_{\Sigma}^2}-1)(d_k^{\dag}d_k+d_kd_k^{\dag}+d_{-k}d_{k}+d_{-k}^{\dag}d_{k}^{\dag})\}\nn
 \\+\{\frac{\omega_{\pi}}{a^3}(b_k^{\dag}b_k+c_kc^{\dag}_k)+\frac{\omega_{\pi}}{2a^3}(\frac{\Omega_{\pi}^2}{\omega_{\pi}^2}-1)(b_k^{\dag}b_k+c_kc_k^{\dag}+b_{-k}c_{k}+c_{-k}^{\dag}b_{k}^{\dag})\}
 \eea

This Hamiltonian has an $su(1,1)$ symmetry and can be diagonalized
by a series of Bogolubov (squeezing) transformations simply given
as follows: In the neutral sector, writing \bea
A_{k}(t,r)&=&\mu(r,t) a_{k}+\nu(r,t)
a^{\dag}_{-k}=U^{-1}(r,t)a_{k}U(r,t) \nn \\
A^{\dag}_{k}(t,r)&=&\nu(r,t) a_{-k}+\mu(r,t)
a^{\dag}_{k}=U^{-1}(r,t)a^{\dag}_{k}U(r,t) \eea
 A similar expansion for diagonalization is done for the sigma
field,with operators $D_{k}(t,r')$ and $D^{\dag}_{k}(t,r')$
similar to the definition of $A_{k}(t,r)$ and $A^{\dag}_{k}(t,r)$
with the d's replacing the a's. For the charged sector \bea
C_{k}(t,r)&=&\mu c_{k} +\nu b^{\dag}_{-k}=U^{-1}(r,t)c_{k}U(r,t)
\nn\\ C^{\dag}_{k}(t,r)&=&\mu c^{\dag}_{k} +\nu
b_{-k}=U^{-1}(r,t)c^{\dag}_{-k}U(r,t)\eea
 \bea
B_{k}(t,r)&=&\mu c_{-k} +\nu b^{\dag}_{k}=U^{-1}(r,t)b_{k}U(r,t)\\
\nn
 B^{\dag}_{k}(t,r)&=&\mu c^{\dag}_{-k} + \nu b_{k}=U^{-1}(r,t)b^{\dag}_{k}U(r,t)\\ \nn \eea
where\be
\mu=Cosh(r)=\sqrt{\frac{1}{2}[(\frac{\Omega_\pi}{\omega_\pi}+\frac{\omega_\pi}{\Omega_\pi})+1]}
\; \;\;\;\;\
\nu=Sinh(r)=\sqrt{\frac{1}{2}[(\frac{\Omega_\pi}{\omega_\pi}+\frac{\omega_\pi}{\Omega_\pi})-1]}
\ee and $r$ is the squeezing parameter, is easily seen to give the
usual result that $\mu^{2}-\nu^{2} = 1$ for a squeezing
transformation.
The complete unitary matrix accomplishing the squeezing transformation may be written down as
\be
U(r,t)=e^{\int \momvol
r(k,t)\{(a^{\dag}_{k}a^{\dag}_{-k}-a_{k}a_{-k})+(d^{\dag}_{k}d^{\dag}_{-k}-d_{k}d_{-k})+(c_{k}b_{-k}+b_{k}c_{-k})-(c^{\dag}_{k}b^{\dag}_{-k}+b^{\dag}_{k}c^{\dag}_{-k})\}}
\ee It should be noted here that putting together all our results
for the neutral and charged sectors, the total diagonalized
Hamiltonian is written in terms of various creation and
annihilation operators as:
\be
H=\int \momvol\frac{1}{2a^3}\{ \Omega_{\pi}\{(A^{\dag}_{k} A_{k} +
\frac{1}{2}) +( C^{\dag}_{k} C_{k} + B^{\dag}_{k}
B_{k}+1)\}+\Omega_{\Sigma}(D^{\dag}_{k}D_{k}+\frac{1}{2})\}. \ee
Since the $\Sigma$ field decouples, we drop all terms associated
with it whenever it is not essential to our arguments allowing us
to write the total dynamical Hamiltonian for the pion fields in
terms of the observed pion creation and annihilation operators
($a,a^{\dag},c,c^{\dag},b$ and $b^{\dag}$): \bea H_{0}&=&\int
\momvol\frac{\Omega_{\pi}}{a^3}\{2(\mu^{2}+\nu^{2})(c^{\dag}_{k}c_{k}
+b^{\dag}_{k}b_{k}+1)
+\mu\nu\{(c_{k}b_{-k}+b_{k}c_{-k})+(c^{\dag}_{k}b^{\dag}_{-k}+b^{\dag}_{k}c^{\dag}_{-k})\}\nn
\\
&+&\frac{\Omega_{\pi}}{a^3}\{(\mu^{2}+\nu^{2})(a^{\dag}_{k}a_{k}+1)+\nu\mu(a^{\dag}_{-k}a^{\dag}_{k}+a_{k}a_{-k})\}
\eea This completes the analysis for the case when $\theta=0$.

\noindent {\bf \underline{CASE 2}}

The second case of interest is when $\rho=\frac{\pi}{2}$. The
Hamiltonian for this case, reduces to the following:
 \bea
 H_{neutral}&=&\int \momvol
\frac{1}{2}\{
\frac{\omega_{\pi}}{a^3}(a_k^{\dag}a_k+a_ka^{\dag}_k)+\frac{a^3v^2\lambda}{2\omega_{\pi}}(1+2Sin^2(\theta))(a_k^{\dag}a_k+a_ka_k^{\dag}+a_{-k}a_{k}+a_{-k}^{\dag}a_{k}^{\dag})
 \nn \\
 &+&\frac{\omega_{\Sigma}}{a^3}(
 d_k^{\dag}d_k+d_kd^{\dag}_k)+\frac{a^3v^2\lambda}{2\omega_{\Sigma}}(1+2Cos^2(\theta))(d_k^{\dag}d_k+d_kd_k^{\dag}+d_{-k}d_{k}+d_{-k}^{\dag}d_{k}^{\dag})\}
 \eea
\be
 H_{charged}=\int \momvol
\{
\frac{\omega_{\pi}}{a^3}(b_k^{\dag}b_k+c_kc^{\dag}_k)+\frac{\omega_{\pi}}{2a^3}(\frac{\Omega_{\pi_\pm}^2}{\omega_{\pi}^2}-1)(b_k^{\dag}b_k+c_kc_k^{\dag}+b_{-k}c_{k}+c_{-k}^{\dag}b_{k}^{\dag})\}
 \ee
 \be
 H_{mixed}=\int \momvol
\{\frac{\lambda
a^3v^2}{2}(\frac{2Sin(\theta)Cos(\theta)}{\sqrt{\omega_\pi\omega_{\Sigma}}})\lrb
 d_k a_{-k}+d_{k}a^{\dag}_k+d^{\dag}_k a_{k}+d^{\dag}_k
 a^{\dag}_{-k}\rrb\}
 \ee
where (for $\rho=\frac{\pi}{2}$),
\bea \frac{\Omega_{\pi}^{2}}{a^6}&=&\frac{k^2}{a^2}+m_{\pi}^2+\lambda
v^2 \nn
\\\frac{\Omega_{\pi_{\pm}}^{2}}{a^6}&=&\frac{k^2}{a^2}+m_{\pi_{\pm}}^2+\lambda
v^2(1+Sin^{2}(\theta) )\eea
 Unlike the case considered above, we now have a non-zero mixing term
 coming from the $\pi_{0}$-$\Sigma$ sector. Since both of these
 are neutral sectors, we can combine the $H_{neutral}$ and
 $H_{mixed}$ terms to form a single neutral sector Hamiltonian which we will
 again call $H_{neutral}$
while $H_{charged}$ remains unchanged. \bea H_{neutral}&=&\int
\momvol\{
\lrb\begin{array}{c}\frac{a_{k}^{\dag}}{\sqrt{\omega_{\pi}}}
\;\;\;\frac{d_{k}^{\dag}}{\sqrt{\omega_{\Sigma}}}\end{array}\rrb\lrb
\begin{array}{c}\frac{a^3}{4}(\frac{\Omega_{\pi}^2+\omega_{\pi}^2)}{a^6}\;\;\;\;\;\;\;\;\;\;\;\;\frac{\lambda
v^2 a^3}{4}Sin(2\theta)\\ \frac{\lambda v^2 a^3}{4}
Sin(2\theta)\;\;\;\;\;\;\;\;\;\;\;\;\frac{a^3}{4}(\frac{\Omega_{\Sigma}^2+\omega_{\Sigma}^2)}{a^6}\end{array}\rrb\lrb\begin{array}{c}\frac{a_{k}}{\sqrt{\omega_{\pi}}}\\
\frac{d_{k}}{\sqrt{\omega_{\Sigma}}}\end{array}\rrb \nn \\
&+&\lrb\begin{array}{c}\frac{a_{k}}{\sqrt{\omega_{\pi}}}
\;\;\;\frac{d_{k}}{\sqrt{\omega_{\Sigma}}}\end{array}\rrb\lrb
\begin{array}{c}\frac{a^3}{4}(\frac{\Omega_{\pi}^2+\omega_{\pi}^2)}{a^6}\;\;\;\;\;\;\;\;\;\;\;\;\frac{\lambda
v^2 a^3}{4}Sin(2\theta)\\ \frac{\lambda v^2 a^3}{4}
Sin(2\theta)\;\;\;\;\;\;\;\;\;\;\;\;\frac{a^3}{4}(\frac{\Omega_{\Sigma}^2+\omega_{\Sigma}^2)}{a^6}\end{array}\rrb\lrb\begin{array}{c}\frac{a^\dag_{k}}{\sqrt{\omega_{\pi}}}\\
\frac{d^\dag_{k}}{\sqrt{\omega_{\Sigma}}}\end{array}\rrb \nn \\
&+&\lrb\begin{array}{c}\frac{a_{-k}}{\sqrt{\omega_{\pi}}}\;\;\;\frac{d_{-k}}{\sqrt{\omega_{\Sigma}}}\end{array}\rrb\lrb\begin{array}{c}\frac{\lambda
v^2a^3}{4}(1+2Sin^2(\theta))\;\;\;\;\;\;\;\;\;\;\;\;\frac{\lambda
v^2a^3}{4}Sin(2\theta)\\\\\frac{\lambda v^2 a^3}{4}
Sin(2\theta)\;\;\;\;\;\;\;\;\;\;\;\;\frac{\lambda
v^2a^3}{4}(1+2Cos^2(\theta))\end{array}\rrb
\lrb\begin{array}{c}\frac{a_{k}}{\sqrt{\omega_{\pi}}}\\\frac{d_{k}}{\sqrt{\omega_{\Sigma}}}\end{array}\rrb\nn
\\&+&\lrb\begin{array}{c}\frac{a_{-k}^{\dag}}{\sqrt{\omega_{\pi}}}\;\;\;\frac{d_{-k}}{\sqrt{\omega_{\Sigma}}}^{\dag}\end{array}\rrb\lrb\begin{array}{c}\frac{\lambda
v^2a^3}{4}(1+2Sin^2(\theta))\;\;\;\;\;\;\;\;\;\;\;\;\frac{\lambda
v^2 a^3}{4}Sin(2\theta)\\\\\frac{\lambda v^2 a^3}{4}
Sin(2\theta)\;\;\;\;\;\;\;\;\;\;\;\;\frac{\lambda v^2
a^3}{4}(1+2Cos^2(\theta))\end{array}\rrb\lrb\begin{array}{c}\frac{a_{k}^{\dag}}{\sqrt{\omega_{\pi}}}\\\frac{d_{k}^{\dag}}{\sqrt{\omega_{\Sigma}}}\end{array}\rrb\}
\eea

These terms clearly show the mixings between the forward and
backward pions as well as the mixing between the $\pi_{0}$ and
$\Sigma$ fields. The misalignment of the vacuum through an angle
$\theta$ induces a mixing of the two fields. The mixed fields are
 \be
\bmat\Bk\\\Ck\emat=\bmat
Cos(\theta)\;\;\;Sin(\theta)\\-Sin(\theta)\;\;\;Cos(\theta)\emat\bmat\ak\\\dk\emat\ee
Then $H_{neutral}$ can be expressed as: \bea
H_{neutral}&=&\bmat\Bdagk\;\;\;\Cdagk\emat\bmat
m_{1}\;\;\;0\\0\;\;\;m_{2}\emat\bmat\Bk\\\Ck\emat
\nn\\&+&\bmat\Bk\;\;\;\Ck\emat\bmat
m_{1}\;\;\;0\\0\;\;\;m_{2}\emat\bmat\Bdagk\\\Cdagk\emat
\nn\\&+&\bmat\B-k\;\;\;\C-k\emat\bmat
n_{1}\;\;\;0\\0\;\;\;n_{2}\emat\bmat\Bk\\\Ck\emat \nn\\&+&\bmat
A^{\dag}_{\theta(-k)}\;\;\; D^{\dag}_{\theta(-k)}\emat\bmat
n_{1}\;\;\;0\\0\;\;\;n_{2}\emat\bmat\Bdagk\\\Cdagk\emat. \eea
Here, \bea 2m_2&=&\frac{\sqrt{\Omega_{\pi}\Omega_{\Sigma}}}{4
a^3}\lrb\sqrt{\frac{\Omega_{\pi}}{\Omega_{\Sigma}}}\lrb\frac{\Omega_{\pi}}{\omega_{\pi}}\lrb1+\frac{1}{2}\sqrt{\frac{\omega_{\pi}}{\omega_{\Sigma}}}\rrb+\frac{\omega_{\pi}}{\Omega_{\pi}}\lrb1-\frac{1}{2}\sqrt{\frac{\omega_{\pi}}{\omega_{\Sigma}}}\rrb\rrb\rrb\nn\\
&+&\frac{\sqrt{\Omega_{\pi}\Omega_{\Sigma}}}{4
a^3}\lrb\sqrt{\frac{\Omega_{\Sigma}}{\Omega_{\pi}}}\lrb\frac{\Omega_{\Sigma}}{\omega_{\Sigma}}\lrb1+\frac{1}{2}\sqrt{\frac{\omega_{\Sigma}}{\omega_{\pi}}}\rrb+\frac{\omega_{\Sigma}}{\Omega_{\Sigma}}\lrb1-\frac{1}{2}\sqrt{\frac{\omega_{\Sigma}}{\omega_{\pi}}}\rrb\rrb\rrb\eea
\bea 2m_1&=&\frac{\sqrt{\Omega_{\pi}\Omega_{\Sigma}}}{4
a^3}\lrb\sqrt{\frac{\Omega_{\pi}}{\Omega_{\Sigma}}}\lrb\frac{\Omega_{\pi}}{\omega_{\pi}}\lrb1-\frac{1}{2}\sqrt{\frac{\omega_{\pi}}{\omega_{\Sigma}}}\rrb+\frac{\omega_{\pi}}{\Omega_{\pi}}\lrb1+\frac{1}{2}\sqrt{\frac{\omega_{\pi}}{\omega_{\Sigma}}}\rrb\rrb\rrb\nn\\
&+&\frac{\sqrt{\Omega_{\pi}\Omega_{\Sigma}}}{4
a^3}\lrb\sqrt{\frac{\Omega_{\Sigma}}{\Omega_{\pi}}}\lrb\frac{\Omega_{\Sigma}}{\omega_{\Sigma}}\lrb1-\frac{1}{2}\sqrt{\frac{\omega_{\Sigma}}{\omega_{\pi}}}\rrb+\frac{\omega_{\Sigma}}{\Omega_{\Sigma}}\lrb1+\frac{1}{2}\sqrt{\frac{\omega_{\Sigma}}{\omega_{\pi}}}\rrb\rrb\rrb\eea
While, \be 2n_2=\frac{\sqrt{\Omega_{\pi}\Omega_{\Sigma}}}{4
a^3}\lrb\sqrt{\frac{\Omega_{\pi}}{\Omega_{\Sigma}}}\lrb\frac{\Omega_{\pi}}{\omega_{\pi}}-\frac{\omega_{\pi}}{\Omega_{\pi}}\rrb\lrb1+\frac{1}{2}\sqrt{\frac{\omega_{\pi}}{\omega_{\Sigma}}}\rrb+\sqrt{\frac{\Omega_{\Sigma}}{\Omega_{\pi}}}\lrb\frac{\Omega_{\Sigma}}{\omega_{\Sigma}}-\frac{\omega_{\Sigma}}{\Omega_{\Sigma}}\rrb\lrb1+\frac{1}{2}\sqrt{\frac{\omega_{\Sigma}}{\omega_{\pi}}}\rrb\rrb\ee
and \be 2n_1=\frac{\sqrt{\Omega_{\pi}\Omega_{\Sigma}}}{4
a^3}\lrb\sqrt{\frac{\Omega_{\pi}}{\Omega_{\Sigma}}}\lrb\frac{\Omega_{\pi}}{\omega_{\pi}}-\frac{\omega_{\pi}}{\Omega_{\pi}}\rrb\lrb1-\frac{1}{2}\sqrt{\frac{\omega_{\pi}}{\omega_{\Sigma}}}\rrb+\sqrt{\frac{\Omega_{\Sigma}}{\Omega_{\pi}}}\lrb\frac{\Omega_{\Sigma}}{\omega_{\Sigma}}-\frac{\omega_{\Sigma}}{\Omega_{\Sigma}}\rrb\lrb1-\frac{1}{2}\sqrt{\frac{\omega_{\Sigma}}{\omega_{\pi}}}\rrb\rrb\ee
The diagonalization procedure for this case along with the
dynamical consequences of this Hamiltonian will be discussed in
\cite{mb2}.

We finally get \bea
H_{neutral}&=&m_{1}\lrb
A^{\dag}_{\theta(k)}A_{\theta(k)}+A_{\theta(k)}A^{\dag}_{\theta(k)}\rrb+n_1\lrb
A_{\theta(k)}A_{\theta(-k)}+A^{\dag}_{\theta(k)}A^{\dag}_{\theta(-k)}\rrb
\nn \\
&+&m_2\lrb
D^{\dag}_{\theta(k)}D_{\theta(k)}+D_{\theta(k)}D^{\dag}_{\theta(k)}\rrb+n_2\lrb
D_{\theta(k)}D_{\theta(-k)}+D^{\dag}_{\theta(k)}D^{\dag}_{\theta(-k)}\rrb\eea
We now apply two squeezing transformations
 \bea
 \Fk&=&\mu A_{\theta(k)}+\nu A^{\dag}_{\theta(-k)} \nn\\
 \Fdagk&=&\mu\Bdagk+\nu\B-k
\eea and \bea \Gk&=&\rho D_{\theta(k)}+\sigma
D^{\dag}_{\theta(-k)}. \nn
\\ \Gdagk&=&\rho D^{\dag}_{\theta(k)}+\sigma D_{\theta(-k)}. \eea Then, \be[\Fk,\Fdagk]=1\ee
and\be[\Gk,\Gdagk]=1\ee implies that the transformations are
indeed Bogolyubov transformations as \be\mu^2-\nu^2=1\ee and
\be\rho^2-\sigma^2=1.\ee The squeezing parameters
$\mu$,$\nu$,$\rho$,$\sigma$ are given by: \bea
2\mu^2&=&\sqrt{\frac{\Omega_{\pi}}{\Omega_{\Sigma}}}\lrb\frac{\Omega_{\pi}}{\omega_{\pi}}\lrb1-\frac{1}{2}\sqrt{\frac{\omega_{\pi}}{\omega_{\Sigma}}}\rrb+\frac{\omega_{\pi}}{\Omega_{\pi}}\lrb1+\frac{1}{2}\sqrt{\frac{\omega_{\pi}}{\omega_{\Sigma}}}\rrb\rrb\nn\\
&+&\sqrt{\frac{\Omega_{\Sigma}}{\Omega_{\pi}}}\lrb\frac{\Omega_{\Sigma}}{\omega_{\Sigma}}\lrb1-\frac{1}{2}\sqrt{\frac{\omega_{\Sigma}}{\omega_{\pi}}}\rrb+\frac{\omega_{\Sigma}}{\Omega_{\Sigma}}\lrb1+\frac{1}{2}\sqrt{\frac{\omega_{\Sigma}}{\omega_{\pi}}}\rrb\rrb\nn\\
&+&1.\eea  \bea
2\nu^2&=&\sqrt{\frac{\Omega_{\pi}}{\Omega_{\Sigma}}}\lrb\frac{\Omega_{\pi}}{\omega_{\pi}}\lrb1-\frac{1}{2}\sqrt{\frac{\omega_{\pi}}{\omega_{\Sigma}}}\rrb+\frac{\omega_{\pi}}{\Omega_{\pi}}\lrb1+\frac{1}{2}\sqrt{\frac{\omega_{\pi}}{\omega_{\Sigma}}}\rrb\rrb\nn\\
&+&\sqrt{\frac{\Omega_{\Sigma}}{\Omega_{\pi}}}\lrb\frac{\Omega_{\Sigma}}{\omega_{\Sigma}}\lrb1-\frac{1}{2}\sqrt{\frac{\omega_{\Sigma}}{\omega_{\pi}}}\rrb+\frac{\omega_{\Sigma}}{\Omega_{\Sigma}}\lrb1+\frac{1}{2}\sqrt{\frac{\omega_{\Sigma}}{\omega_{\pi}}}\rrb\rrb\nn\\
&-&1.\eea \bea
2\rho^2&=&\sqrt{\frac{\Omega_{\pi}}{\Omega_{\Sigma}}}\lrb\frac{\Omega_{\pi}}{\omega_{\pi}}\lrb1+\frac{1}{2}\sqrt{\frac{\omega_{\pi}}{\omega_{\Sigma}}}\rrb+\frac{\omega_{\pi}}{\Omega_{\pi}}\lrb1-\frac{1}{2}\sqrt{\frac{\omega_{\pi}}{\omega_{\Sigma}}}\rrb\rrb\nn\\
&+&\sqrt{\frac{\Omega_{\Sigma}}{\Omega_{\pi}}}\lrb\frac{\Omega_{\Sigma}}{\omega_{\Sigma}}\lrb1+\frac{1}{2}\sqrt{\frac{\omega_{\Sigma}}{\omega_{\pi}}}\rrb+\frac{\omega_{\Sigma}}{\Omega_{\Sigma}}\lrb1-\frac{1}{2}\sqrt{\frac{\omega_{\Sigma}}{\omega_{\pi}}}\rrb\rrb\nn\\
&+&1.\eea \bea
2\sigma^2&=&\sqrt{\frac{\Omega_{\pi}}{\Omega_{\Sigma}}}\lrb\frac{\Omega_{\pi}}{\omega_{\pi}}\lrb1+\frac{1}{2}\sqrt{\frac{\omega_{\pi}}{\omega_{\Sigma}}}\rrb+\frac{\omega_{\pi}}{\Omega_{\pi}}\lrb1-\frac{1}{2}\sqrt{\frac{\omega_{\pi}}{\omega_{\Sigma}}}\rrb\rrb\nn\\
&+&\sqrt{\frac{\Omega_{\Sigma}}{\Omega_{\pi}}}\lrb\frac{\Omega_{\Sigma}}{\omega_{\Sigma}}\lrb1+\frac{1}{2}\sqrt{\frac{\omega_{\Sigma}}{\omega_{\pi}}}\rrb+\frac{\omega_{\Sigma}}{\Omega_{\Sigma}}\lrb1-\frac{1}{2}\sqrt{\frac{\omega_{\Sigma}}{\omega_{\pi}}}\rrb\rrb\nn\\
&-&1.\eea With these definitions the neutral sector  Hamiltonian
is simply:
\be
H_{neutral}=\frac{\sqrt{\Omega_{\pi}\Omega_{\Sigma}}}{4
a^3}\lrb\lrb\Fdagk\Fk+\frac{1}{2}\rrb+\lrb\Gdagk\Gk+\frac{1}{2}\rrb\rrb\ee

Combining with the charged sector
\be
H_{charged}=\frac{\Omega_{\pi_\pm}}{a^3}\{ C^{\dag}_{k} C_{k} +
B^{\dag}_{k} B_{k}+1\} \ee the total Hamiltonian for the case
$\rho =\frac{\pi}{2}$

\be H_{\pi/2}=\int \momvol \{ \frac{\Omega_{\pi_\pm}}{a^3}(
C^{\dag}_{k} C_{k} +
B^{\dag}_{k}B_{k}+1)+\frac{\sqrt{\Omega_{\pi}\Omega_{\Sigma}}}{4
a^3}\lrb(\Fdagk\Fk+\frac{1}{2})+(\Gdagk\Gk+\frac{1}{2})\rrb\}\ee
with \bea \frac{\Omega_{\pi}^{2}}{a^6}&=&\frac{k^2}{a^2}+m_{\pi}^2+\lambda
v^2 \nn
\\\frac{\Omega_{\pi_{\pm}}^{2}}{a^6}&=&\frac{k^2}{a^2}+m_{\pi_{\pm}}^2+\lambda
v^2(1+Sin^{2}(\theta) )\eea

This completes our analysis of the quantization of the Hamiltonian for the two cases mentioned above.
We have taken an $O(4)$ sigma model and succeeded in quantizing the quadratic fluctuations to arrive at two Hamiltonians which provide all the required ingredients for analysing the formation, evolution and eventual decay of the DCC.
Indeed, in case 2, the Hamiltonian shows explicit mixing with an angle ($\theta$), providing a mixing
 parameter and therefore a misalignment parameter in isospin space.

\section{Evolution of the fluctuations and parametric amplifications}
In the last section we have constructed the quantum Hamiltonians
for two cases ( $\theta=0$ and $\rho=\frac{\pi}{2}$ )of the $O(4)
$ sigma model. We have also shown the diagonalization of the
Hamiltonians so that they can be written in terms of the
appropriate quantum fluctuation fields as purely quadratic
Hamiltonians. In this section we shall explicitly consider the
case when $\bf{\theta=0} $ (the dynamics of the case
$\rho=\frac{\pi}{2}$ will be given in a subsequent paper
\cite{mb2}). We notice that H has the form of a decoupled
Hamiltonian. This is easy to understand from the $S0(4)$ parent.
The S0(4) vector has been decomposed into four fields:
$\pi_{\pm}$,$\pi_0$ and $\Sigma$ being respectively the charged
pions, the neutral pion and the sigma fields.

It is the sigma field which decouples in this particular
Hamiltonian and therefore it can be analyzed independently of the
pion fields. As our interest is in the pion fields,  we can write
the total dynamical Hamiltonian for the pion fields in terms of
the observed pion creation and annihilation operators
($a,a^{\dag},c,c^{\dag},b $ and $b^{\dag}$)in terms of the
squeezing parameters as: \bea H&=&\int
\momvol\frac{\Omega_{\pi}}{a^3}\{2(\mu^{2}+\nu^{2})\{c^{\dag}_{k}c_{k}
+b^{\dag}_{k}b_{k}+1\}+\mu\nu\{(c_{k}b_{-k}+b_{k}c_{-k})
+(c^{\dag}_{k}b^{\dag}_{-k}+b^{\dag}_{k}c^{\dag}_{-k})\} \nn
\\+&&\frac{\Omega_{\pi}}{a^3}\{(\mu^{2}+\nu^{2})\{a^{\dag}_{k}a_{k}+1\}+(\nu\mu)\{a^{\dag}_{-k}a^{\dag}_{k}+a_{k}a_{-k}\}\}
\eea

We now define  the bilinear operators \bea
 {{{\cal{D}}}}&=&a_{k}a_{-k}+b_{k}c_{-k}+c_{k}b_{-k}=K_1^- +K_2^-
 +K_3^-
 \nn\\
 {{{\cal{D}}}}^{\dagger}&=&a^{\dag}_{-k}a^{\dag}_{k}+c^{\dag}_{-k}b^{\dag}_{k}+b^{\dag}_{-k}c^{\dag}_{k}=K_1^+ +K_2^+
 +K_3^+
 \nn\\
 N&=&\frac{1}{2}\{a^{\dag}_{k}a_{k}+a^{\dag}_{-k}a_{-k}+b^{\dag}_{k}b_{k}+b^{\dag}_{-k}b_{-k}+c^{\dag}_{k}c_{k}+c^{\dag}_{-k}c_{-k}+3\}=K_1^0+K_2^0+K_3^0
 \lb{bl}\eea and it is easy to see that they satisfy an $su(1,1)$ algebra \be
[N,{{{\cal{D}}}}]=-{{{\cal{D}}}};\;\;\;\;
[N,{{{\cal{D}}}}^{\dag}]={{{\cal{D}}}}^{\dag};\;\;\;
[{{{\cal{D}}}}^{\dag},{{{\cal{D}}}}]=-2N. \ee The $su(1,1)$
invariant Hamiltonian for the pion fields assumes the form
\begin{equation}
H=\int \momvol\frac{1}{a^3}
2\Omega_{\pi}(k,t)(\mu^2+\nu^{2})N+2\Omega_{_\pi}(k,t)\mu\nu({{\cal{D}}}+{{\cal{D}}}^{\dag})
\end{equation}

The  time dependent evolution equation for the eigenstates of H is
given by
\be
H(t)|\psi(t)>=i\frac{d}{dt}|\psi(t)> \ee The particular $su(1,1)$
structure elucidated above provides us the solution:
 \be |\psi(t)>=e^{\int \momvol
r_k({{\cal{D}}}^{\dag}_{k}-{{\cal{D}}}_{k})}|\psi(0> \lb{wf} \ee
for the evolution of the wave function immediately \cite{perel}.
Here $r_k$ is the squeezing parameter related to the physical
variables $\Omega_{\pi}(k,t)$ and $\omega_{\pi}(k)$ through \be
Tanh(2r_k)= \frac{(\frac{\Omega_{\pi}(k,t)}{\omega_\pi})^2-1}
 {(\frac{\Omega_{\pi}(k,t)}{\omega_\pi})^2+1} \ee Where $\Omega_{\pi}(k,t\longrightarrow\infty)=\omega_{\pi}(k)$.
Thus in the evolution of the condensate , it is the  frequency
changes which  bring about squeezing \cite{Lo}.

The diagonalized Hamiltonian $H_{0}$ can be converted into a Hamiltonian in terms of quantum fields corresponding to
the operators $A,B,C $ and their adjoints to obtain a purely quadratic Hamiltonian
 (the starting point of many early works on the subject of the
 DCC)\cite{mina,bm}.

We can, for example, write \be
\frac{\Omega_\pi}{a^3}(\Adagk\Ak+\frac{1}{2})=\frac{\Omega_\pi}{a^3}
(\Adagk\Ak
+\Ak\Adagk)=(\frac{\Omega_\pi}{a^3})^2\Pi_A^2(k,t)+P_{\Pi_{A}}^2(k,t).\ee
Similarly for B and C, we write: \bea
\frac{\Omega_\pi}{a^3}(B^{\dag}_kB_k+\frac{1}{2})=\frac{\Omega_\pi}{a^3}
(B^\dag_kB_k+B_kB^{\dag}_k)=(\frac{\Omega_\pi}{a^3})^2\Pi_B^2(k,t)+P_{\Pi_{B}}^2(k,t)\nn
\\
\frac{\Omega_\pi}{a^3}(C^\dag_kC_k+\frac{1}{2})=\frac{\Omega_\pi}{a^3}
(C^\dag_kC_k+C_kC^\dag_k)=(\frac{\Omega_\pi}{a^3})^2\Pi_C^2(k,t)+P_{\Pi_{C}}^2(k,t)
\\ \eea
The Hamiltonian $H_0$ can then be written as:
\be H_0(t)=\int\frac{d^3k}{(2\pi)^3}\sum_{i=A,B,C}\frac{1}{2}((\frac{\Omega_{\pi}}{a^3})^2\Pi_{i}^2(k,t)+P_{\Pi_{i}}^2(k,t))\ee

The Schroedinger equation for each momentum mode is simply: \be
H_0(k,t)\psi(k,t)=i\frac{d}{dt}\psi(k,t).\ee

If we use the $\Pi$-representation (co-ordinate space
representation) for $\psi(k,t)$, then, the $su(1,1)$ symmetry of
the Hamiltonian tells us that the solution for $\psi(k,t)$ is just
a Gaussian.  For simplicity we work with a Gaussian of the form
\begin{equation}
<\pi_0,\pi_{+},\pi_{-}|\psi>(t)=\prod_{k,i}L_i(t)exp^{(-W_i(t)\Pi_{i}^2)}
\ee while the complete wavefunction is: \be
\psi(t)=\int\frac{d^3k}{(2\pi)^3}\psi(k,t).\ee

Then for each mode A,B,C,D the wave function $\psi_{k,i}(t)$
evolves as
\begin{equation}
i\frac{\partial\psi_{k,i}(t)}{\partial
t}=(i\frac{\dot{L_i}}{L_i}-i\Pi_{i}^2(k)\dot{W}_{i}(k,t))\psi_{k,i}(t)
\end{equation}
While
\be
H_{k,i}\psi_{k,i}(t)=\frac{1}{2a^3}\{\Omega_{\pi_i}^2\Pi_i^2(k,t)-\frac{\partial^2\psi_{k,i}(t)}{\partial
\Pi_i^2} \}. \ee Combining the above we find that $ W_i(t)$ and
$\psi_{k,i}(t)$ evolve as:
\begin{equation}
W_i(t)=-ia^3/2 \frac{\dot{\psi}_{k,i}(t)}{\psi_{k,i}(t)}
\end{equation}
\be
\ddot{\psi_{i,k}}(t)+3\frac{\dot{a}}{a}\dot{\psi_{i,k}}(t)-\frac{\Omega_{\pi_i}^2}{a^3}\psi_{i,k}(t)=0.
\ee

The equation satisfied by the wavefunctions for each mode are then
given by: \bea
\ddot\psi_{A}(k,t)+\frac{3\dot{a}}{a}\dot{\psi_{A}}+(\frac{\Omega_{\pi}}{a^3})^{2}(k,t)\psi_{A}(k,t)=0\nn
\\
\ddot\psi_{B}(k,t)+\frac{3\dot{a}}{a}\dot{\psi_{B}}+(\frac{\Omega_{\pi}}{a^3})^{2}(k,t)\psi_{B}(k,t)=0\nn
\\
\ddot\psi_{C}(k,t)+\frac{3\dot{a}}{a}\dot{\psi_{C}}+(\frac{\Omega_{\pi}}{a^3})^{2}(k,t)\psi_{C}(k,t)=0\eea
where \be
(\frac{\Omega_\pi}{a^3})^2(k,t)=(\frac{\veck^2}{a^2})+m_{\pi}^2+\lambda
v^2\ee

Where it may be recalled that  $A_{k}(t),B_{k}(t)$ and $C_{k}(t)$
are related to $a_{k},b_{k}$ and $c_{k}$ (the physical pion
operators), by the squeezing transformation given by \bea
A_{k}(t,r)&=&U^{-1}(r,t)a_{k}U(r,t)
\\B_{k}(t,r)&=&U^{-1}(r,t)b_{k}U(r,t)\\ \nn
C_{k}(t,r)&=&U^{-1}(r,t)c_{k}U(r,t)\eea

$Cosh(r)=\sqrt{\frac{1}{2}[(\frac{\Omega_\pi}{\omega_\pi}+\frac{\omega_\pi}{\Omega_\pi})+1]}
\; \;\;\;\;\
\nu=Sinh(r)=\sqrt{\frac{1}{2}[(\frac{\Omega_\pi}{\omega_\pi}+\frac{\omega_\pi}{\Omega_\pi})-1]}$
and $r$ is the squeezing parameter. \be U(r,t)=e^{\int \momvol
r(k,t)((a^{\dag}_{k}a^{\dag}_{-k}-a_{k}a_{-k})+d^{\dag}_{k}d^{\dag}_{-k}-d_{k}d_{-k})
+(c_{k}b_{-k}+b_{k}c_{-k})-(c^{\dag}_{k}b^{\dag}_{-k}+b^{\dag}_{k}c^{\dag}_{-k}))}
\lb{wf1} \ee

The expectation values of the number operator for the neutral
pions for each momentum k is given by: \be
<\psi_{k}(t)|a^{\dag}_{k}a_{k}|\psi_{k}(t)>=Sinh^{2}(r)=<\psi_{k}|A^{\dag}_{k}(t)A_{k}(t)|\psi_{k}>.\ee
A similar expression may be obtained for the charged pions.

Since we are considering the expansion of the plasma in a
spherically symmetric manner as emphasized by the Robertson-Walker
type metric $ds^2=dt^2-a(t)^2 d\vecx^2$, it is possible to scale
the time so as to provide a conformally flat metric: we let \be
d\eta=a(t)^{-1}dt\ee so that \be
ds^2=a(\eta)^2(d\eta^2-d\vecx^2).\ee The equations of motion given
above can be transformed into ones that resemble a harmonic
oscillator with time dependent frequencies. We shall write only
the generic form of the above equations: In terms of the scaled
time, $\eta$, we have: \be \psi''+\frac{2 a'}{a}\psi' +(\vec
k^2+(m_{\pi}^2+{\lambda} (<\Phi^2>-f_{\pi}^2)
  )a^2\psi =0\ee
where a prime denotes differential wrt $\eta$. Here we have a few
remarks. Firstly we note that to make contact with the dynamical
equations for the pions used as starting points for analysing the
DCC, we return to the point about the potential that we have
chosen vs the traditional potential chosen in such studies. The
traditional potential is $(\Phi_{i}^{2}-f_{\pi}^{2})^{2}$ while we
have taken a potential $ (\Phi_{i}^{2})^{2}$. Therefore, the
relationship between our work and the early studies is
accomplished through a replacement
$v^{2}\longrightarrow(<\Phi^{2}>(\eta)-f_{\pi}^{2})$. This
explains the equation written above and the further analysis of
this paper. Lastly, let us scale $\psi$: \be \xi=a\psi \ee so that
the equation becomes:
\be
-\xi''+V(\eta)\xi =(k^2+m_{\pi}^2)\xi \ee where \be
V(\eta)=a^{-1}\frac{d^{2}a}{d\eta^2}+m_{\pi}^{2}(1-a^{2})-\lambda
(<\Phi^2>-f_{\pi}^2).\lb{dual}\ee Thus in the symmetry broken
stage $<\Phi^2>=f_{\pi}^2$
\be
V_b(\eta)=a^{-1}\frac{d^{2}a}{d\eta^2}+m_{\pi}^{2}(1-a^{2})
\lb{vb}\ee And in the symmetry restored stage$<\Phi^2>=0$
\be
V_r(\eta)=a^{-1}\frac{d^{2}a}{d\eta^2}+m_{\pi}^{2}(1-a^{2})+\lambda
a^{2}f_{\pi}^2. \lb{vr}\ee

 These equations have a dual personality: on the one hand they
are Schroedinger like equations with $\eta$ corresponding to the
"spatial" like variable and $E=\omega_{\pi}^2$. Therefore, they
allow calculation of the reflection and transmission coefficients
over the ``potential barrier'' provided by the $V(\eta)$ term. On
the other hand, they can also be looked upon as equations for time
dependent harmonic oscillators with time dependent frequencies
given by $\Omega_{\pi}^2$ and $\omega_{\pi}^2$. These two pictures
enable us to calculate the squeezing parameter dependent number
operator $N(k)=Sinh^2(r(k))$ involved in the evolution of the
plasma as can be seen in appendix 1.

We also note that the expansion coefficient $a(\eta)$ provides us
with a control parameter on the expansion rate of the plasma while
the system provides us with another parameter, which we call
$\tau$  for the ``rolling down'' of the fields in the potential.
According to the pioneering papers \cite{bj,kw}, when the rate of
expansion is greater then the relaxation rate of the field the DCC
forms.

We now consider two cases: the first being a toy model
\cite{Amado} without the effects of expansion and  the second
being a more realistic "Baked Alaska Situation".

In the first case, let us suppose that $a(\eta)=1$ so that
\be
V(\eta)=-\lambda (<\Phi^2>-f_{\pi}^2). \ee Then the equation for
$\xi$ is:\\
 \be
\xi''+(k^2+m_{\pi}^2+\lambda (<\Phi>^2-f_{\pi}^2))\xi =0. \ee

If $\tau$ is the time the state spends  in the symmetry restored
phase, a quench can be modelled by assuming that for
$-\frac{\tau}{2}\le\eta\le\frac{\tau}{2}$ the vacuum expectation
value $<\Phi^2>=0$ and for $\eta
>\frac{\tau}{2}$ the vacuum relaxes to its value
$<\Phi^2>=<v^2>\approx f_{\pi}^2$ ( since the potential is
translationally invariant, this could easily be mapped to the
interval $\{0,\tau \}$, but our choice makes the potential
symmetric) . For this sudden quench the problem reduces to that of
transmission through a symmetric rectangular potential barrier of
height $(\lambda f_{\pi}^2)=\frac{m_{\Sigma}^{2}}{2}$ and width
$\tau$.

The transmission coefficient for such a barrier is easily
calculable  \cite{terhaar,Flugge}and in terms of physically known
values is:

\be
T=\frac{1}{1+(\frac{(\frac{m_{\Sigma}^2}{2})^{2}}{( \omega_k^2)
(\frac{m_{\Sigma}^2}{2}-\omega_k^2)})Sinh^2[(\frac{m_{\Sigma}^2}{2}-\omega_k^2)^{\frac{1}{2}}\tau]}
.\ee  From appendix 1 we get
\be
N(k)=\frac{1-T}{T}=\frac{(\frac{m_{\Sigma}^2}{2})^2}{ \omega_k^2
(\frac{m_{\Sigma}^2}{2}-\omega_k^2)}Sinh^2[(\frac{m_{\Sigma}^2}{2}-\omega_k^2)^{\frac{1}{2}}\tau]
\ee

 The dependence of N(k) on k for different values of $\tau$ is shown in fig. 1,
 clearly exhibiting the amplification of the low momentum modes.

\begin{figure}[htbp]
\begin{center}
\epsfxsize=10cm\epsfbox{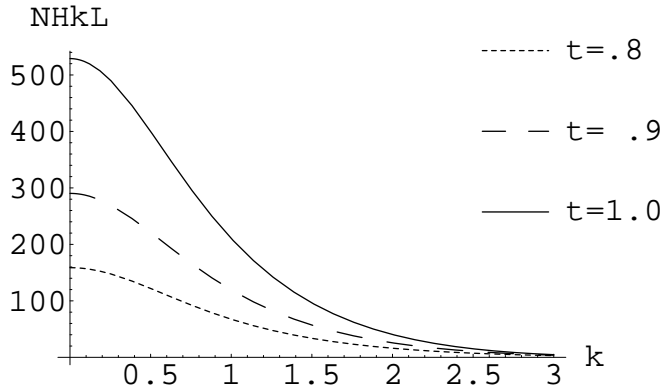} \end{center} \caption{Shows
the variation of $N(k)$ and $k$( in units of $m_{\pi}$) with
$\tau=.8,.9,1.0$ in units of $m_{\pi}^{-1}$ for the quenched limit
}
\end{figure}

Figure 1 also shows us that the longer the system stays in the
state of broken symmetry the larger the DCC domains. The
dependence of N(k) on $\tau$ for different values of $k$ is shown
in fig. 2. Recall that by definition, the amplification of the
zero modes constitutes DCC formation. Since we have a
"Schrodinger" wave equation that is exactly solvable,  we can also
calculate the size of the DCC domains. We shall leave that for a
subsequent paper \cite{mb2}.

\begin{figure}[htbp]
\begin{center}
\epsfxsize=10cm\epsfbox{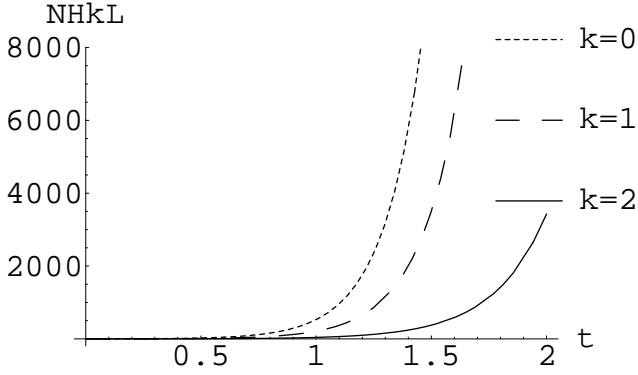} \end{center} \caption{Shows
the variation of $N(k)$ with $\tau$(units of $m_{\pi}^-1)$ for
$k=0,1$ and $2$ (units of $m_{\pi}$) for the quenched limit }
\end{figure}
 This case is
similar to that considered by ref \cite{Amado}.

For a more realistic scenario, the expansion must be included to
show  that the enhancement of the low energy modes and the
squeezing parameter are dependent on the rate of the expansion
mechanism by which the symmetry is restored.
 To produce substantial squeezing, we require a quenched scenario.
 To show this we have to compare the situation
of a sudden quench with a slow adiabatic relaxation of the system
from the symmetry restored stage  to the symmetry broken stage.
The transition between quenching and adiabaticity can be modeled
in two ways in view of the dual nature of eqn(\ref{dual}).

We consider  the expansion coefficient $a(\eta)$ to be of the form
\be
a^2(\eta)=\Theta(-\eta)Tanh[-\frac{b}{2}(\eta+\frac{\tau}{2})]
+\Theta(\eta)Tanh[\frac{b}{2}(\eta-\frac{\tau}{2})]. \lb{at}\ee
Here $\frac{1}{b}$ measures the rate of the expansion and we
choose b to be in units of $m_{\pi}$. When $b\tau>>1$ we have the
quenched limit (fast expansion)  and  enhancement of low momentum
modes should occur (DCC formation). When $b\tau<<1$ we have the
adiabatic limit and no enhancement of low momentum modes should
occur (no DCC formation).

When viewed as a time dependent oscillator  equation in an
expanding metric we may write the equation for $\xi$ in the
symmetry restored phase as an oscillator equation
\be
\xi''+\omega_{\pi}^2\xi-V_r(\eta)\xi =0 \ee and in the symmetry
broken stage as \be \xi''+\omega_{\pi}^2\xi-V_b(\eta)\xi =0.\ee
Where
 $V_b(\eta)$and $V_r(\eta)$ are given by (\ref{vr},\ref{vb}) with the particular choice
of the expansion parameter $a(\eta)$ given by (\ref{at}). Thus the
change in frequency from the restored to the broken stage is given
by
\be
V_r(\eta)-V_b(\eta)=+\frac{m_{\Sigma}^2}{2}a(\eta)^2 \ee

Given that we have to calculate the transmission coefficients we
need  to look at the wave functions $\xi$ in the limit $\eta->\pm
\infty$. This is possible by solving the oscillator equation with
a variable frequency:\be
\xi''+\omega_{\pi}^2+\frac{m_{\Sigma}^2}{2}[1-a^2(\eta)]\xi=0.
\lb{os}\ee We see that the above equation(\ref{os}) satisfies the
required limits $\Omega(\eta)^2=\omega_{\pi}^2$ as $\eta->\pm
\infty$ and between these limits it gradually goes to a maximum at
$\Omega(\eta)=\Omega_{\pi}^2$.
 The equation (\ref{os}) can be converted into a  Schrodinger wave equation for
a  Wood -Saxon potential barrier given by

\be
V(\eta)=V_0
\lrb\Theta(-\eta)(\frac{1}{1+e^{-b(\eta+\frac{\tau}{2})}})+\Theta(\eta)(\frac{1}{1+e^{b(\eta-\frac{\tau}{2})}})
.\rrb\ee Where $V_0=\frac{m_{\Sigma}^2}{2}.$  The comparison of
the rectangular potential barrier with this barrier is shown in
fig.3, which reveals it to be a good approximation in the
adiabatic limit.

\begin{figure}[htbp]
\begin{center}
\epsfxsize=10cm\epsfbox{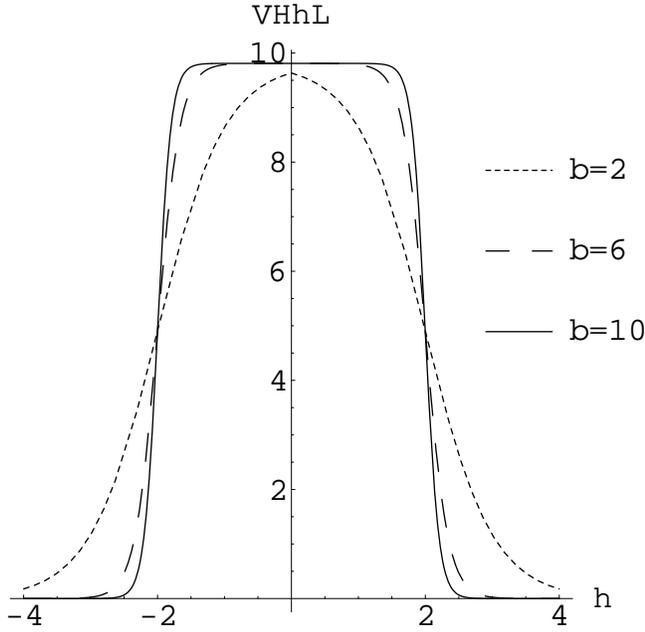} \end{center} \caption{Shows the
variation of the Wood Saxon Potential Barrier of width $\tau=4$
with b. Large $ b\tau$  gives the Rectangular Potential Barrier
(quenched limit) }
\end{figure}
We take the values of the variables
$E=(k^2+\omega_{\pi}^2)^{\frac{1}{2}}$  and $k'^2=E-V_{0}$.
 The transmission coefficient
for this barrier is given by:
\be
T_{ws}=\frac{Sinh^2(\pi\frac{2k}{b})Sinh^2(\pi\frac{2k'}{b})}{Sinh^4(\pi\frac{(k-k')}{b})\{
 4|C|Sin^2(k'b)+(|C|^2-1)\}}\;\;; E\ge V_{0}
 \ee
 and the same with $k'--->ik'$ for $E<V_{0}$
 where
 $C=\frac{Sinh^2(\pi\frac{2(k+k')}{b})}{Sinh^2(\pi\frac{2(k-k')}{b})}$.\\
Note that for b large this reduces exactly to the rectangular
potential barrier (quenched limit) transmission coefficient  and
for b very small this goes over to the Poschl-Teller (Eckart)
Barrier transmission coefficient.

The number of particles of mode $k$ equals .
\be
N(k)=\frac{1-(Sinh^4(\pi\frac{(k-k')}{b})\{
 4|C|Sin^2(k'b)+(|C|^2-1)\})}{Sinh^2(\pi\frac{2k}{b})Sinh^2(\pi\frac{2k'}{b})} \ee
 Here b
measures the duration of the quench. $N(k) vs. k $ is plotted in
fig. 4 . We see that in the adiabatic limit of small $b\tau$ ,
$N(k)$ is exponentially suppressed so that there is no enhancement
of low momentum modes.

\begin{figure}[htbp]
\begin{center}
\epsfxsize=10cm\epsfbox{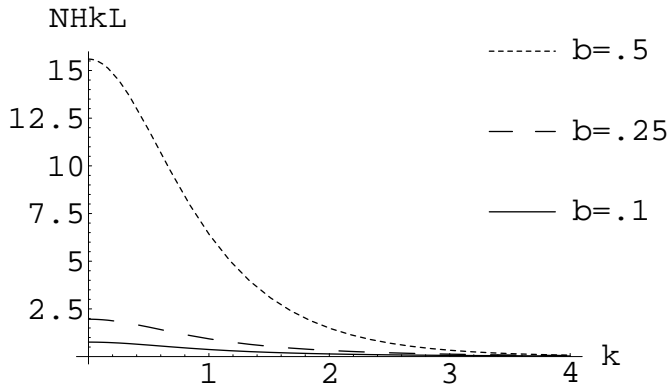} \end{center} \caption{Shows the
variation N(k) with k for values of  b in the {\it adiabatic
limit} for Wood- Saxon barrier . }
\end{figure}
Figure 5 shows the variation of $N(k)$ with $k$ for large values
of b , showing the enhancement of low momentum modes in the
quenched limit.
\begin{figure}[htbp]
\begin{center}
\epsfxsize=10cm\epsfbox{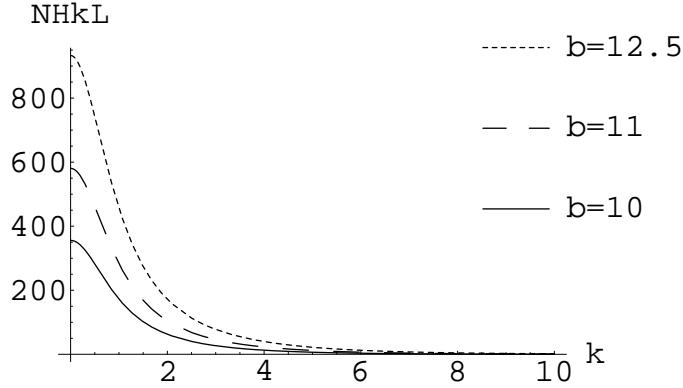} \end{center} \caption{Shows
the variation N(k) with k for values of  b in the {\bf quenched
limit} for Wood- Saxon barrier  . }
\end{figure}

From the above we conclude that if expansion  time
($\frac{1}{b}$)is faster then the rolling time ($\tau$) we get the
quenched limit, while, if expansion time  is slower then the
rolling time we get the adiabatic limit. Since in the squeezed
state description $N(k)=Sinh^2r_k$ ,where $r_k=r(k)$ is the
squeezing parameter , we conclude that in the quenched limit the
squeezing parameter is much greater than in the adiabatic limit.
This demonstrates clearly the connection between the rate of
expansion and squeezing and the formation of the DCC. If squeezing
is large we have formed a DCC , if not then there is no DCC. This
will now enable us to give signatures of DCC formation which are
related to the formation process of the DCC.

\section{Pion Radiation from DCC}
Having shown that the dynamics of the  evolution of the  DCC
suggests a squeezed state treatment of emerging pion waves and
that this effect is more pronounced in quenched scenario than an
adiabatic one, we now proceed to show the possible experimental
signals that would result. A complete treatment, with the
incorporation of isospin and disorientation in isospin space for
arbirary momenta is given in \cite{mb1} and \cite{mb2}. For the
present  however, since we see an amplification of the low
momentum modes we consider only the case when $k=0$. In such case
for pions near zero momentum $k\longrightarrow 0$ the state
(\ref{wf})  factors into a squeezed state for the neutral pions
and a Caves- Schumaker state for the charged pions. In this limit
the bilinear operators of equation (\ref{bl})  are \bea
 {{{\cal{D}}}}&=&
 a_0^2+b_{0}c_{0}+c_{0}b_{0}\\ \nn
 {{{\cal{D}}}}^{\dagger}&=&a_0^{\dag 2}+c_{0}^{\dag}b_0^{\dag}+b_{0}^{\dag}c_0^{\dag}\\ \nn
 N&=& a^{\dag}_{0}a_{0}++b^{\dag}_{0}b_{0}+c^{\dag}_{0}c_{0}+3/2,
\eea

Thus the k=0 wave function is

\be
|\psi>=e^{r_0({\cal {D}}^{\dag}-{\cal {D}})}|\psi_0(0)>=
e^{r_0(a_0^{\dag^2}+c^{\dag}_{0}b^{\dag}_{0}+b^{\dag}_{0}c^{\dag}_{0}-a_0^2+b_{0}c_{0}+c_{0}b_{0})}|\psi_0(0)>
\ee The pion multiplicity distribution is given by
\be
P_{n_0,n_+,n_-}=|<n_{0},n_{+},n_{-}|\psi>|^{2}=
<n_{0}|e^{r_0(a_{0}^{\dag})^{2}-r_0*a_{0}^{2}}|0>
<n_{+},n_{-}|e^{2r_0 (b_{0} ^{\dag}c_{0}^{\dag}-b_{0}c_{0})}|0>|^{2} \ee
defining $S(r_0)$ as the one mode squeezing operator
\begin{eqnarray}
S(r_0)&=&
<n_{0}|e^{r_0 ((a_{0}^{\dag})^{2}-a_{0}^{2})}|0>
\nonumber \\ &=& S_{n_{0},0}.
\end{eqnarray}
$S^{tm}(r_0)$ is then the two mode squeezing operator
\begin{eqnarray}
<n_{+},n_{-}|e^{(r_0(b_{0}^{\dag}c_{0}^{\dag}-b_{0}c_{0}))}|0> \nonumber \\
=S^{tm}_{n_{+},n_{-},0}.
\end{eqnarray}
The neutral and charged pion distribution is:
\begin{equation}
P_{n_0,n_c}=<S_{n_{0},0}>^{2}<S^{\dag}{}^{m}_{n_{+},n_{-},0,0}>^{2}
\end{equation}
which is just the product of squeezed distributions for charged
and neutral pions and only even number of pions emerge. Writing
$n_{+}=n_{-}=n_{c}$, we get the distribution of charged particles
to be
\begin{equation}
P_{n_{c}}{}=\sum_{n_0}P_{n_0,n_c}=
\frac{(tanh(r_0))^{2n_{c}}}{(cosh(r_0))^{2}}
\end{equation}
and
\begin{equation}
P_{n_{0}}=\sum_{n_c}P_{n_0,n_c}=\frac{n_{0}!(tanh(r_0))^{n_{0}}}{((\frac{n_{0}}{2})!)^{2}cosh(r_0)
2^{n_{0}}}
\end{equation}

The generalized squeezed  eigenstate leads to products of two
types of squeezed states of pions at zero momentum. The neutral
pions being in a one mode squeezed state and the charged pions
being in an SU(1,1) coherent or two-mode squeezed state. Thus the
neutral and charged pion distributions are significantly different
as the two types of states have different properties. We now
illustrate the effect of squeezing in these two distributions.
Figs. 6, and 7 show the difference in the charged and neutral pion
distributions as we vary from the squeezing parameter  from a low
value to  a high value.

\begin{figure}[htbp]
\begin{center}
\epsfxsize=10cm\epsfbox{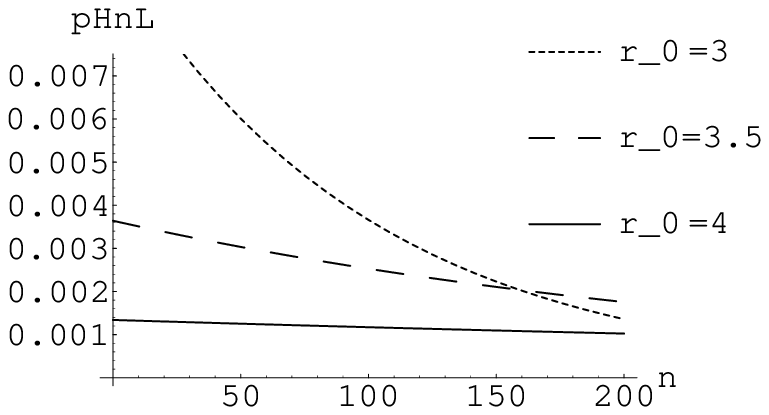} \end{center} \caption{Shows the
variation of $P_{nc}$ with $n$ for the $r_{0}=3,3.5$ and $ 4$}
\end{figure}

\begin{figure}[htbp]
\begin{center}
\epsfxsize=10cm\epsfbox{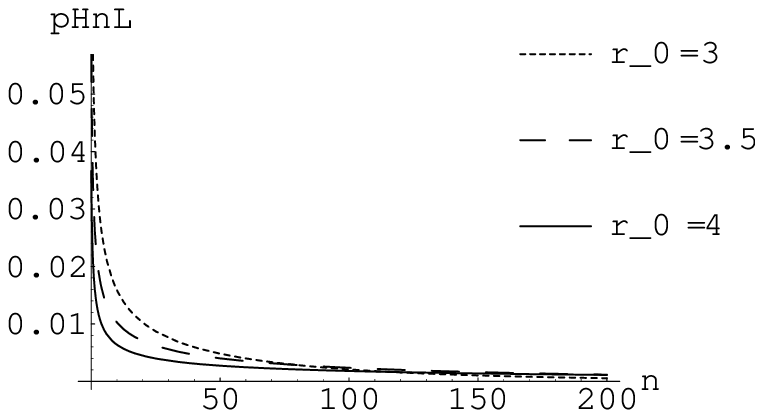} \end{center} \caption{Shows the
variation of $P_{nc}$ with $n$ for the $r_{0}=3,3.5$ and $ 4$}
\end{figure}

  We now illustrate the effect of quenching versus
adiabaticity on these two distributions. Figs. 8  shows the
difference in the charged and neutral pion distribution for as we
vary from the adiabatic limit where the difference is negligible
to the quenched limit where the difference is significant.
\begin{figure}[htbp]
\begin{center}
\epsfxsize=10cm\epsfbox{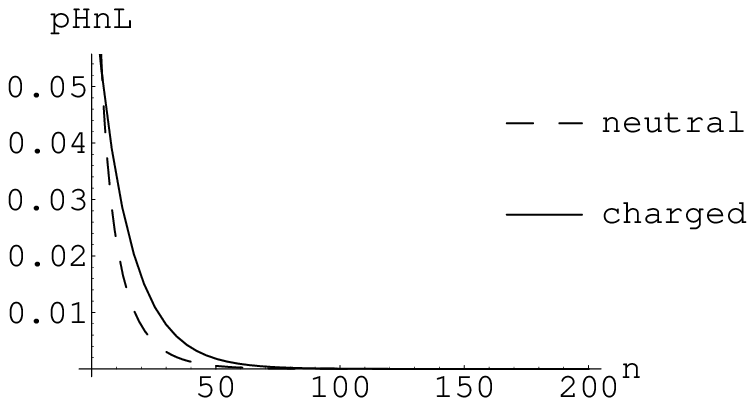}\end{center} \caption{Shows the
variation of $P_{n0}$ (solid line)and $P_{nc}$(dashed line) with
$n$ for the {\it adiabatic} limit ($r_0=2$)}
\end{figure}

\begin{figure}[htbp]
\begin{center}
\epsfxsize=10cm\epsfbox{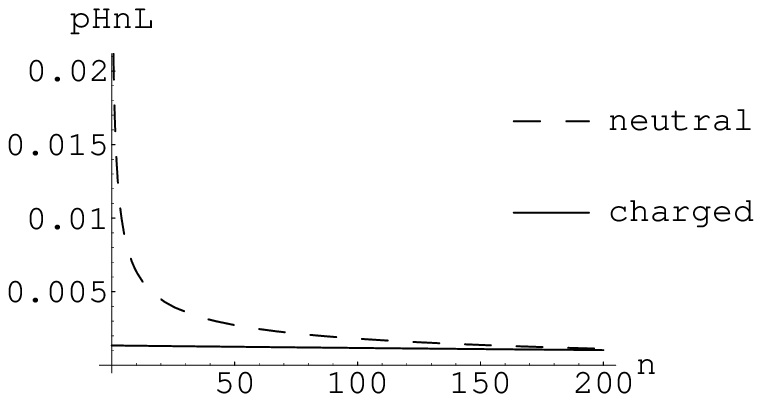}\end{center} \caption{Shows the
variation of $P_{n0}$ (solid line)and $P_{nc}$(dashed line) with
$n$ for the {\bf quenched} limit ($r_0=4$)}
\end{figure}

We now calculate the correlation function by first calculating the
the generating function corresponding to the multiplicity
distribution $P_n$. It is given by $Q(\lambda)=\sum_\lambda
(1-\lambda)^n P_n$. The two particle zero momentum correlation
function $G^2(0)$ is given by \be
G^2(0)=\frac{<n^2>-<n>}{<n^2>}=\frac{\frac{\partial^2Q}{\partial
\lambda^2}|_{\lambda=1}}{( \frac{\partial Q}{\partial
\lambda})^2|_{\lambda=1}}\ee
 the variation of $G^2_{neutral}\over G^2_{charged}$ with  the squeezing parameter
 is given in  figure 10.
\begin{center}
\begin{figure}[htbp]
\epsfxsize=10cm\epsfbox{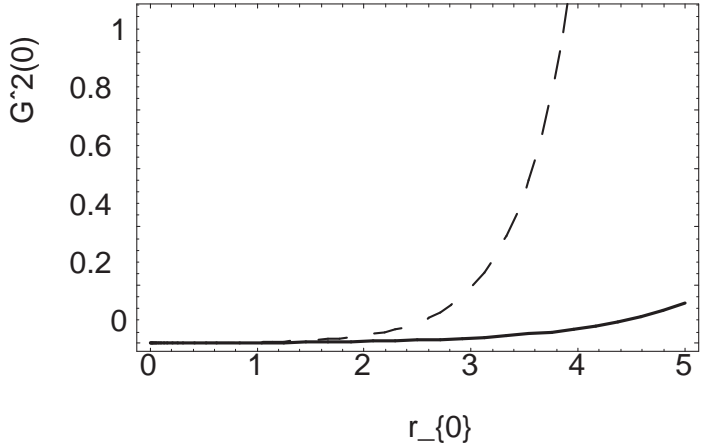} \caption{Shows the difference in
correlations of the charged(dashed line) and neutral pions(solid
line) as a function of the squeezing parameter $r_{0}$}
\end{figure}
\end{center}

Thus we see that in the adiabatic limit  $G^2_{neutral}\approx
G^2_{charged}$, whereas in the quenched limit $G^2_{neutral}<<<
G^2_{charged}$, giving a very clear indication of the effect of
DCC formation on the Bose -Einstein correlations.

To conclude this section, we have shown that the sudden quench
approximation in the evolution of the disoriented chiral
condensate leads to a substantial amount of squeezing which
manifests itself in the dramatic difference between charged and
neutral pion distributions. For an adiabatic quench the difference
is much less, so that both the total multiplicity distributions of
charged and neutral pions as well as their second order
correlation functions are dramatic characteristic signals for the
DCC  and are related directly to the way in which the DCC forms.
These are unambiguous, therefore they must be examined thoroughly
in searches for the DCC \cite{Mohanty}.

\section{Conclusion.}

To conclude, in this paper we have constructed effective
Hamiltonians  for the evolution of the disoriented chiral
condensate without  and with orientation in isospin space,
starting from an O(4) sigma model through the inclusion of second
order quantum fluctuations. We have shown that both Hamiltonians
have SU(1,1) symmetries leading to the presence of squeezed staes
in their dynamics.  Unlike most earlier studies, our calculations
of the effective Hamiltonians are not restricted to zero momentum
and take care of back to back momentum correlations. The evolution
of the wave function in an expanding metric, for one case (without
orientation in isospin space) has been considered in detail. The
competing effects of the expansion rate and the rolling down time
of the system from a state of restored to broken symmetry has been
explicitly examined. We find that in the quenched limit ( fast
expansion) the low momentum modes are enhanced signaling DCC
formation, whereas in the adiabatic (slow expansion) limit, no
such enhancement occurs. This has been shown to be directly
related to the value of the squeezing parameter. The manifestation
of this difference shows up directly in the total neutral and
charged multiplicity distributions at zero momentum and the second
order correlation functions.

Further work is required to incorporate the effects of isospin.
This is being done in a  subsequent publication \cite{mb1}.
Furthermore, the evolution of the wave function corresponding to
the Hamiltonian with orientation in isospin space is also the
subject of a forthcoming communication \cite{mb2}. Within this
framework back to back momentum correlations and event by event
analysis of the experimental signals can also be done. This will
provide a solid unified picture of the various models of DCC
formation , evolution and decay, together with new experimental
signals. Finally, we have achieved our goal of answering Bjorken's
DCC troubling questions posed at the Trento meeting \cite{bj3}.
\begin{itemize}
\item{Question a}: Are coherent states the right
quantum DCC description, or should one go beyond to squeezed
states, etc.
\item{Answer a}: YES, one is naturally lead to squeezed states
from a quantum field theoretic perspective.
\item{Question b}: How does one link DCC thinking
to quantum effects in the data, especially the Bose­Einstein
correlations. Is DCC just another way of talking about the same
thing?
\item{Answer b}: DCC formation leads to a significant difference
between the charged and neutral pion Bose-Einstein correlations
corresponding to large squeezing. So, it is not another way of
talking about the same thing.
\item{Question c}: How does one enforce quantum number
conservation, especially charge?
\item{Answer c}: Charge conservation is automatically guaranteed
by  the isospin analysis shown in paper \cite{mb1}.
\item{Question d}: DCC production may imply anomalous bremsstrahlung, due to
the large quantum fluctuations in charge. Can this be calculated
from first principles?
\item{Answer d}: This can be easily incorporated in our model and
will be shown in a later communication.
\item {Question e}: Can one really set up the problem
at the quantum field theory level?
\item{ Answer e}: An emphatic YES! Albeit, the full quantum
theoretical effective Hamiltonian with the contribution of the
quarks still remains to be considered.
\end{itemize}
\section{Acknowledgements.}
B.A.B. would like to thank IUCAA, Pune, India, for its support
under the {\it Senior Associate Scheme}. She would also like to
thank Prof. Tapan Nayak for his interest in this work. C.M. would
like to thank Prof.K.N. Pathak, Vice chancellor, Panjab University
and Prof.A.K. Kapoor, University of Hyderabad, for facilitating
this collaboration. C.M's research was actively hindered by the
University Grants Commission (India) under its ``Research
Scientist'' scheme.  As a result, his research was fully funded
from his salary as a UGC (India) ``Research Scientist C (Professor
level!!)''.

\section{Appendix 1:
Relationship between the transmission coefficients and squeezing
parameter for a time dependent harmonic oscillator} The process of
particle creation and the excitation of parametric oscillators can
be related to transmission and reflection of shrodinger waves over
a potential barrier. The method we describe for the relation is
closely related to ref \cite{HU}
 Let us consider the
evolution of the wave function of the equation \be
\xi''_{i,k}+\lrb k^2-V_{i,k}(\eta)\rrb\xi_{i,k}=0\ee if we define
$k^2-V_{i,k}(\eta)=\omega(\eta)^2$, then  the above equation is a
time dependent harmonic oscillator
$\xi''(t)+\omega^2(\eta)\xi(t)=0$ . In the event of a change in
frequency of the oscillator  from $\omega_-$ to $\omega_+$
 the asymptotic form of the real
solution at $\eta\longrightarrow\pm\infty$ is
\be
\xi_{\pm}(t)=\frac{1}{2}(a_{\pm}e^{i\omega_{\pm}t}+a_{\pm}^{*}e^{-\omega_{\pm}t})\ee

We compare this with the complex solution  of a one-d oscillator
treated as a reflection over a barrier $k^2-V_{i,k}$ with $\eta$
corresponding to the spatial variable of a Shrodinger type of
equation :

 \bea \xi_{c-}(t)&=&e^{i\omega_{-}t}+Re^{-i\omega_{-}t} \nn
\\ \xi_{c+}(t)&=&Te^{i\omega_{+}t} \eea
To identify
 $a_-=1+R^*$ and $a_+=T$\\.

Now consider the particle creation problem modelled by a time
dependent harmonic oscillator \cite{Lo}. The solution can be
represented by
 \bea
\xi_{r-}&=&\frac{e^{-i\omega_i t}}{\sqrt{2\omega_i}}\\ \nn
 \xi_{r+}(t)&=&\alpha
\frac{e^{-i\omega_f t}}{\sqrt{2\omega_f t}}+\frac{\beta
e^{i\omega_{f}t}}{\sqrt{2 \omega_f t}}\eea

Now with the definition of the transmission coefficient
\be
T=\frac{|\xi_{out}|^2v_{out}}{|\xi_{in}|^2v_in} \ee

  If we
regard t as a spatial variable then these represent reflection and
transmission over a one-d barrier. In particle creation problems
 the
potential barrier reflection and the transmission coefficient can
be related to the squeezing parameter in a particle creation
problem by the relation $Sinh^2r=\frac{R}{T}=<n>=\beta \beta^{*}$
\end{document}